\documentclass[aps,prb, longbibliography, showpacs,twocolumn,floats,epsfig,pdflatex]{revtex4}
\usepackage{amssymb}
\usepackage{amsbsy}
\usepackage{amsmath}
\usepackage{epsfig}
\usepackage{amsfonts}
\usepackage{graphicx}
\usepackage{amssymb}
\usepackage{upgreek}
\usepackage{dsfont}
\usepackage{bm}
\usepackage{xcolor, comment}
\usepackage{hyperref}

\def\grad{\boldsymbol\nabla}
\def\Om{{\boldsymbol\Omega}_{\bf p}}

\begin{document}

%\title {Chiral anomaly and plasma oscillations in the transient photovoltage of Weyl semimetals}
%\title {Photoinduced plasma oscillations in Weyl semimetals}
\title {Van Roosbroeck's equations with topological terms: the case of Weyl semimetals}

\author{Pierre-Antoine Graham}
\author{Simon Bertrand}
\author{Michaël Bédard}
\author{Robin Durand}
\author{Ion Garate}

\affiliation{D\'epartement de Physique, Institut Quantique and Regroupement Qu\'eb\'ecois sur les Mat\'eriaux de Pointe, Universit\'e de Sherbrooke, Sherbrooke, Qu\'ebec, Canada J1K 2R1}
\date{\today}
\begin{abstract}

Van Roosbroeck's equations constitute a versatile tool to determine the dynamics of electrons under time- and space-dependent perturbations. 
Extensively utilized in ordinary semiconductors, their potential to model devices made from topological materials remains untapped.
Here, we adapt van Roosbroeck's equations to theoretically study the bulk response of a Weyl semimetal to an ultrafast and spatially localized light pulse in the presence of a quantizing magnetic field. 
We predict a transient oscillatory photovoltage that originates from the chiral anomaly. The oscillations take place at the plasma frequency (THz range) and are damped by intervalley scattering and dielectric relaxation.
Our results illustrate the ability of van Roosbroeck's equations to unveil the interplay between electronic band topology and fast carrier dynamics in microelectronic devices. 

\end{abstract}
\maketitle

%{\bf Review SM section references}

\section{Introduction}

Van Roosbroeck's (VR) system of equations\cite{Roosbroeck1950} comprises the drift-diffusion and continuity equations for electric charge carriers with the Poisson equation for the electric field.
These equations have been key for the understanding and development of landmark devices, such as transistors, solar cells and photodiodes. 
As such, numerous studies and monographs have been published about the VR equations and their solutions (see e.g. Refs. [\onlinecite{Demari1968, Fonstad1994, McKelvey1984, Selberherr1984, farrell2016numerical, Sze2021}]). 
Modern commercial software packages\cite{multiphysics2016semiconductor} are also equipped to solve the VR equations in a variety of realistic device geometries.
Yet, the aforementioned works and software have been tailored to topologically trivial materials.

%To date, commercial microelectronic devices have been based on materials with trivial electronic band topology. Yet, 
It is now known that a large fraction of solids host topologically nontrivial electrons \cite{Vergniory2019,Wieder2021, Vergniory2022}.
Thus, there is a marked interest towards creating devices that will exploit the topological properties of matter \cite{Liu2020, Gilbert2021}. 
Remarkably, although drift-diffusion equations have been used to predict novel electrical transport properties in certain topological materials under restricted (e.g. static) conditions  \cite{Parameswaran2014}, full-fledged VR equations remain vastly underexploited therein.
Little is known about fundamental changes that could emerge in the solutions of those equations when electrons have a nontrivial band topology. If topological microelectronic devices are to become technological reality, such gap of knowledge must be filled.
The objective of our paper is to make progress in this direction and to show that VR equations, appropriately adapted to account for nontrivial electronic topology,  can unveil new physical effects in topological devices of potential technological interest.

The family of topological materials being large and diverse, we are inclined to make a choice for the purposes of the present study. The material that we focus on is a Weyl semimetal (WSM), where pairs of nondegenerate electronic bands cross at isolated points in the Brillouin zone [\onlinecite{Armitage2018}]. These points, called Weyl nodes, 
are sources or sinks of Berry curvature and have a chirality index $\chi =\pm 1$, which is a topological invariant of the electronic band structure. 

\begin{figure}[t]
  \begin{center}
    \includegraphics[width=0.8\columnwidth]{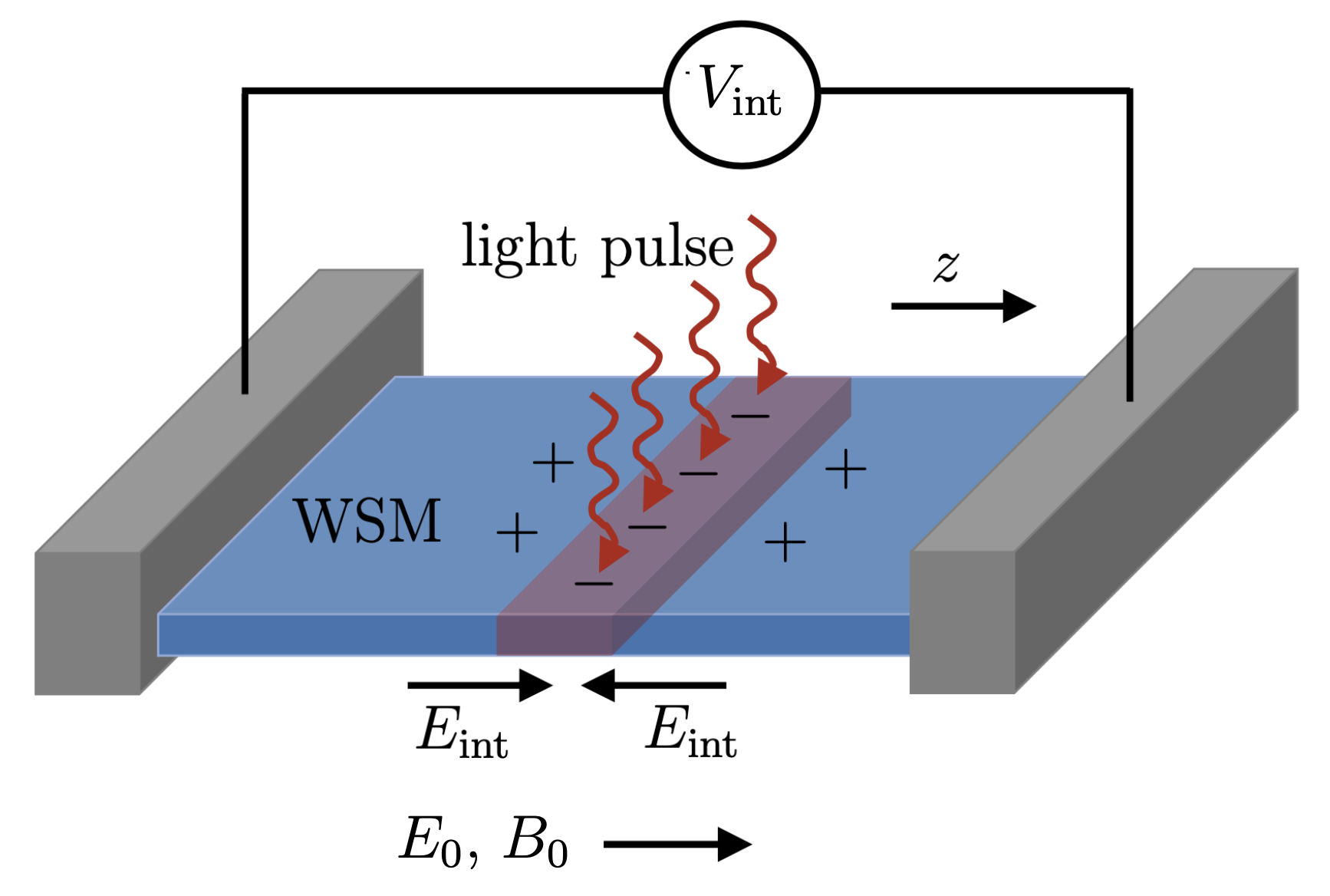}                                                                                                
\caption{A Weyl semimetal film (blue) is placed between two contacts (grey). Initially, the WSM is subjected to an electric field $E_0\hat{\bf z}$ and a magnetic field $B_0\hat{\bf z}$, both uniform and static. Then, a finite region of length $dz$ (red) is illuminated with a light pulse of duration $dt$. 
Photoexcited electrons ($-$) and holes ($+$) drift and diffuse away unequally from the illuminated region, 
%At the outset, the holes move faster than the electrons, 
giving rise to a local charge accumulation and a concomitant electric field $E_{\rm int}$. The line integral of $E_{\rm int}$ between the contacts yields a transient photovoltage $V_{\rm int}$ that oscillates at the plasma frequency due to the chiral anomaly.}
 \label{fig:setup}
  \end{center}
\end{figure} 

Recently, the investigation and control of WSM using ultrafast light has emerged as a frontier of fundamental and applied research \cite{Weber2021, Bao2021}. On one hand, the influence of the Berry curvature in photodetection and nonlinear optics has been highlighted \cite{Golub2018,Anh2020, Liu2020, Ma2021}.
On the other hand, experiments \cite{Jadidi2020, Levy2020, Cheng2021} have measured the light-induced chiral anomaly,
a topological property whereby collinear electric and magnetic fields induce a transfer of electrons between Weyl nodes of opposite chirality \cite{Burkov2015}. 

In this paper, we investigate the interplay between electronic band topology and transient carrier dynamics in a WSM irradiated by a spatially inhomogeneous light pulse (Fig.~\ref{fig:setup}),  using a new approach.
We begin by writing VR equations for a WSM placed under a strong magnetic field  and subject to a light pulse (Sec.~\ref{sec:basic}), with adjustments to accommodate for the nontrivial electronic band topology.
We follow in Secs.~\ref{sec:lin_main} and \ref{sec:photo} by linearizing and solving the preceding VR equations.  
Our strategy of solution is to posit simple but physically justified boundary conditions, and then to take advantage of them by integrating the VR equations over the system's length. 
This approach enables us to gain analytical insight for a physical quantity of interest, namely the photovoltage.
We thus find a transient photovoltage that oscillates at the plasma frequency. The oscillations originate from the chiral anomaly and are driven by an internal electric field that results from the spatial separation between photoexcited electrons and holes. Unlike in Refs.~[\onlinecite{Jadidi2020, Levy2020, Cheng2021}], the effect of the chiral anomaly is present even when the electric field of light is perpendicular to the static magnetic field.
The main text of the paper ends with some discussion (Sec.~\ref{sec:disc}) and conclusions (Sec.~\ref{sec:conc}).
% with a brief summary and a list of open questions.
The six appendices contain technical aspects that allow to reproduce the main results of the paper.

\section{Basic equations}
\label{sec:basic}

\begin{figure}[t]
  \begin{center}
    \includegraphics[width=0.8\columnwidth]{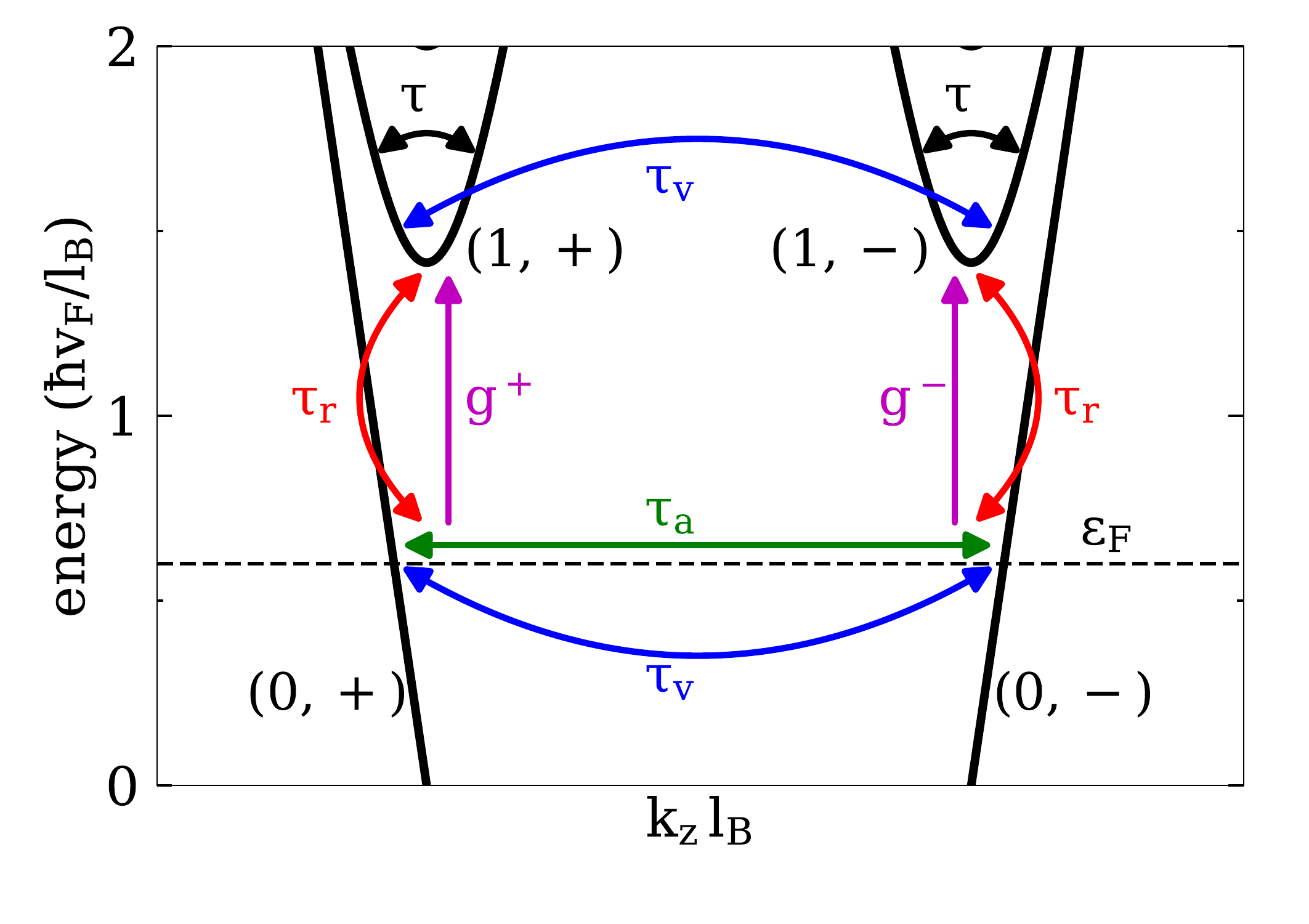}                                                                                               
\caption{Low-energy Landau bands $(n, \chi)$, with $n=0,1$ and $\chi=\pm 1$. 
The $n=0$ bands are unidirectional, due to the nontrivial electronic band topology. 
The magnetic length is $l_B$ and $k_z$ stands for the electronic wave vector component parallel to the static magnetic field.  The equilibrium Fermi energy is $\epsilon_F$ (dashed lines). Other important parameters are (i) the intraband relaxation time $\tau$ in the $n=1$ bands, (ii) the intervalley relaxation time $\tau_v$, (iii) the intravalley electron-hole recombination time $\tau_r$, (iv) the characteristic time $\tau_a$  for intervalley charge transfer along chiral Landau levels due to the chiral anomaly, (v) the optical generation rates $g^\pm$. In the main text, we adopt the hierarchy  $\tau<\tau_a \lesssim\tau_r\ll\tau_v$.}
 \label{fig:ql}
  \end{center}
\end{figure}

In this section, we adapt the VR equations to a bulk WSM with two nodes.
The nodes of opposite chirality are related by symmetry and the energy dispersion is untilted.
A strong static and uniform magnetic field ${\bf B}_0=B_0\hat{\bf z}$ is applied, so that the equilibrium Fermi energy $\epsilon_F$ intersects solely with the chiral ($n=0$) Landau levels (Fig.~\ref{fig:ql}).
It is in this regime that the effect of chiral anomaly in VR equations is most pronounced.
In App. ~\ref{sec:weak_field}, we present the corresponding theory for a weak $B_0$.
 
We restrict our analysis to the electronic dynamics in the $n=0$ and $n=1$ Landau levels.
We assume that (i) all the other bands are far enough from $\epsilon_F$, and their electronic populations unchanged by the light pulse;
(ii) the occupation of the $n=1$ bands in thermal equilibrium is small (low temperature);
(iii) 
 the light pulse is uniform across the film thickness and width, but nonuniform along the film length $z$ (Fig.~\ref{fig:setup}). 
Assumption (iii) justifies the use of one-dimensional VR equations, which we enumerate and discuss next.

In the drift-diffusion approximation, the charge current density along $z$ for the band $(1,\chi)$
is
\begin{equation}
\label{eq:j1}
j_1^\chi = q \mu_1 \rho_1^\chi E + q D_1 \partial_z\rho_1^\chi,
\end{equation}
where $\rho_1^\chi$ is the electron number density in band $(1,\chi)$, $q$ is the absolute value of the electron's charge ($q>0$), $E$ is the $z$ component of the total electric field (including a static and uniform electric field $E_0$, the electric field of light $E_{\rm light}$, and an internal electric field $E_{\rm int}$ discussed below), 
$\mu_1 =q \tau/m^*=q |v| \tau l_B/(\sqrt{2}\hbar)$
is the mobility of electrons in the $n=1$ band in the effective mass ($m^*$) approximation, $|v|$ is the Fermi velocity, $\tau$ is the intraband scattering time in band $(1,\chi)$, $l_B = \sqrt{\hbar/(q |B_0|)}$ is the magnetic length and $D_1=v^2\tau$ is the diffusion coefficient.
Hereafter, $\tau$ is assumed to be the shortest of all characteristic times in the problem, thereby justifying time-locality 
 in Eq.~(\ref{eq:j1}).

The charge current density along $z$ in band $(0,\chi)$ is
\begin{equation}
\label{eq:j0}
j_0^\chi = \chi q v \rho_0^\chi,
\end{equation}
where $v= |v| {\rm sgn}(B_0)$ is the (constant) slope of band $(0,-)$ and $\rho_0^\chi$ is the electron number density therein.
The reason why Eq.~(\ref{eq:j0}) looks different from Eq.~(\ref{eq:j1}) is that the motion of electrons in a chiral Landau level is one-way.
Accordingly, one cannot write the current in a {\em single} chiral Landau level as a sum of drift and diffusion currents.
This is an example of how electronic band topology requires adjusting the VR equations away from their traditional form.
%In Ref.~[\onlinecite{sm2022}] (Sec. A), we show that 
It turns out that the total current in the chiral Landau levels, $j_0^+ + j_0^-$, can be written in terms of drift and diffusion currents if time variations of the current are slow on the scale of the intervalley relaxation time $\tau_v$ (see App. ~\ref{sec:drift_diffusion}).
Since we are interested in the dynamics that is faster than $\tau_v$, $j_0^+ + j_0^-$ is not simply a sum of drift and diffusion currents.

The charge continuity equation for band $(n,\chi)$ reads
\begin{equation}
\label{eq:cont}
\partial_z j_n^\chi - q \partial_t\rho_n^\chi = q R_n^\chi+ q G_n^\chi - \chi \frac{q^3 E B_0}{4\pi^2\hbar^2}  \delta_{n,0},
\end{equation}
where
\begin{equation}
\label{eq:gen1}
G_0^\chi = - G_1^\chi \equiv g^\chi(z,t)/2
\end{equation}
is the $(0,\chi)\to (1,\chi)$ optical generation rate ($g^\chi>0$) in units of $1/(\text{time}\times\text{volume})$, 
\begin{align}
\label{eq:rel1}
R_1^\chi&= \chi\frac{\rho_1^+-\rho_1^-}{\tau_v} + \frac{\rho_1^\chi - \rho_{1,\rm eq}^\chi}{\tau_r}\\
\label{eq:rel0}
R_0^\chi &= \chi\frac{\rho_0^+-\rho_0^-}{\tau_v} - \frac{\rho_1^\chi - \rho_{1,\rm eq}^\chi}{\tau_r}  %- \chi \frac{q^2 E B_0}{4\pi^2\hbar^2} 
\end{align}
are the relaxation rates for the excess charge in the relaxation time approximation, 
$\tau_r$ is the intravalley electron-hole recombination time and $\rho_{1,\rm eq}^\chi$ is the equilibrium electron density in band $(1,\chi)$.
We have for simplicity assumed that $\tau_v$ is the same for $n=0$ and $n=1$. 
By considering all relaxation times to be constants, we focus on electron dynamics not far from equilibrium.

The second term in the right hand side (r.h.s.) of Eq.~(\ref{eq:rel0}) ensures the conservation of the total charge
via
$\sum_{n, \chi} \left( \partial_z j_n^\chi - q \partial_t\rho_n^\chi\right)=0.$
The third term in the r.h.s. of Eq.~(\ref{eq:cont}) is the chiral anomaly term, which in the low-temperature regime enters directly only in the continuity equation for the band intersecting the Fermi level ($n=0$)
\footnote{We neglect the magnetic field of the light pulse and we likewise neglect light-induced lattice strains and their possible contributions \cite{ilan2020pseudo} to the chiral anomaly term.}.
This term, absent in traditional VR equations, is another example of the adaptation required by nontrivial band topology.
%We neglect this possibility, which deserves further study.} 

In Eq.~(\ref{eq:gen1}), $g^\chi$ 
%is regarded as a phenomenological function of $z$ and $t$, though it 
is calculable from Fermi's golden rule~(see App.~\ref{sec:fermi}).  
Two properties of $g^\chi$ worth noting are that (i) $g^\chi\propto |B_0|$ due to the Landau level degeneracy, and (ii) $g^+(z,t) = g^-(z,t)$
if the light pulse preserves the symmetry relating the two Weyl nodes.
%we expect $g^+(z,t) = g^-(z,t)$.
The latter property fails when $E_0\neq 0$, which breaks the $z\to -z$ symmetry.
Before light illumination, Eqs.~(\ref{eq:cont}), (\ref{eq:rel1}) and (\ref{eq:rel0}) yield
%$E_0$ causes a steady-state imbalance between $\rho_0^+$ and $\rho_0^-$, through the interplay between chiral anomaly and the intervalley relaxation. 
%Solving Eqs.~(\ref{eq:cont}), (\ref{eq:rel1}) and (\ref{eq:rel0}) in the absence of light, we have
\begin{equation}
\label{eq:nolight}
\rho_0^\chi=  \rho_{0,\rm eq}^\chi +\chi \frac{q^2 \tau_v}{8\pi^2 \hbar^2} E_0 B_0 \text{    and  } \rho_1^\chi = \rho_{1,\rm eq}^\chi,
\end{equation}
which implies $\rho_0^+-\rho_0^-\propto E_0  \neq 0$, as though the two Weyl nodes have different chemical potentials. 
If so, $g^+\neq g^-$ even when the light pulse preserves the crystal symmetry.

For later reference, the electric current in the absence of a light pulse is obtained from Eq.~(\ref{eq:nolight}) and reads 
\begin{align}
\label{eq:j_nolight}
j 
&= \sigma_0 E_0 + q  v_d (\rho_{1,\rm eq}^+ + \rho_{1,\rm eq}^-),
\end{align}
where $v_d=\mu_1 E_0$ is the drift velocity in $n=1$ bands and
\begin{equation}
\label{eq:sigma0}
\sigma_0= q^3 v B_0 \tau_v/(4\pi^2\hbar^2)
\end{equation}
is the dc conductivity from $n=0$ bands.

Lastly, the longitudinal part of the electric field obeys the Poisson equation
\begin{equation}
\label{eq:poisson}
\partial_z E = -(q/\epsilon) \sum_{n=0,1} \sum_{\chi=\pm 1} \left(\rho_n^\chi-\rho_{n,\rm eq}^\chi\right),
\end{equation}
where $\epsilon$ is the dielectric constant of the material and the equilibrium charge densities are assumed to be spatially uniform (homogeneous doping).

Because $E_{\rm light}$ is transverse and $E_0$ is independent of $z$,
$E$ in Eq.~(\ref{eq:poisson}) equals an internal, photoinduced electric field $E_{\rm int}$.
%, which originates from the light pulse. 
Initially, photoexcited electrons and holes propagate at different speeds due to the markedly different energy dispersions of the $n=0$ and $n=1$ bands.
Consequently, a local net charge is generated in the region where the light pulse acts, which in turn produces $E_{\rm int}$.
This field tends to neutralize the local charge at times exceeding the dielectric relaxation time. 

Neglecting thermoelectric effects \cite{Massicotte2021} for simplicity, Eqs. (\ref{eq:j1})-(\ref{eq:rel0}) and (\ref{eq:poisson}) form the complete VR system of equations for the unknowns  $\rho_n^\chi$ and $E$.
These nonlinear and coupled equations must in general be solved numerically.
Nevertheless, as we show next, analytical insights about the transient optoelectronic response of WSM can be gained by linearizing the VR equations and solving them with simple boundary conditions. 

\section{Linearized equations}
\label{sec:lin_main}

Let us define $\Sigma_n\equiv \sum_\chi \left(\rho_n^\chi-\rho_{n,\rm eq}^\chi\right)$ and $\Delta_n \equiv \sum_\chi \chi\left(\rho_n^\chi-\rho_{n,\rm eq}^\chi\right)$ as the deviations of the scalar and chiral electron densities from equilibrium (respectively).
For an optical pulse of modest intensity, VR equations can be linearized in $\Sigma_n$ and $\Delta_n$ (see App.~\ref{sec:linearized}):
% the outcome reads
\begin{align}
\label{eq:vr1}
&\left(D_1 \partial^2_z+v_d \partial_z - \partial_t-1/\tau_r\right)\Sigma_1 = -g_s \nonumber\\
&\left( D_1 \partial_z^2+ v_d\partial_z- \partial_t-1/\tau_r-2/\tau_v\right)\Delta_1 =-g_d\nonumber\\
&v\partial_z \Delta_0 - \partial_t \Sigma_0 +\Sigma_1/\tau_r=  g_s\nonumber\\
&v \partial_z \Sigma_0 - \left(\partial_t+2/\tau_v \right)\Delta_0 + \Delta_1/\tau_r= g_d- \epsilon E/(q v \tau_a^2)  \nonumber\\
&\Sigma_0 + \Sigma_1 = -\epsilon \partial_z E_{\rm int}/q,
\end{align}
where 
$g_s = \sum_\chi g^\chi/2$ and $g_d = \sum_\chi \chi g^\chi/2$
are the scalar and chiral optical generation rates (respectively), and
\begin{equation}
\label{eq:tau_a}
%\frac{1}{\tau_a^2} \equiv \frac{q^3 v B}{2\pi^2\hbar^2\epsilon} = \frac{2}{\tau_v}\frac{\sigma_0}{\epsilon} 
\tau_a = \sqrt{\frac{\tau_v}{2}\frac{\epsilon}{\sigma_0}}
\end{equation}
is a characteristic time that emerges from the chiral anomaly term in Eq.~(\ref{eq:cont}). 
Physically,  $\epsilon/\sigma_0$ is the dielectric relaxation time and $\tau_a^{-1}$ is the bulk plasma frequency of chiral electrons (the zero of the $zz$ component of the dielectric tensor) in the absence of scattering \cite{Parent2020}.
% for the excess of charge in the chiral Landau levels.

In Eq.~(\ref{eq:vr1}), the electron dynamics in the $n=0$ bands is affected by the dynamics in the $n=1$ bands, but not vice versa.
%It is known  that $E_{\rm light}$ can lead to a dynamical internodal charge pumping by virtue of the chiral anomaly.
Also, the fourth line of Eq.~(\ref{eq:vr1}) captures the dynamical internodal charge pumping induced by $E_{\rm light}$ by virtue of the chiral anomaly \cite{Jadidi2020,  Levy2020, Cheng2021}.
%Such effect is captured by the third line of Eq.~(\ref{eq:vr1}). 
Hereafter, we assume that $E_{\rm light}=0$ (i.e., ${\bf E}_{\rm light} \perp {\bf B}_0$).
Even then, $E_{\rm int}$ drives a dynamical chiral anomaly because it is collinear to $B_0$. Next, we investigate its physical consequence.

\section{Transient photovoltage}
\label{sec:photo}

Equation~(\ref{eq:vr1}) can be solved analytically for a WSM film of length $L$ placed between two contacts, and subject to a light pulse that has a finite extent in space and time (Fig.~\ref{fig:setup}). The light pulse is centered at time $t=0$ and the spatial region where it acts is sufficiently far from the contacts.
Accordingly, the deviations of the carrier densities with respect to equilibrium vanish when $z\to\pm L/2$ (contact location) or $t\to\pm\infty$.
These boundary conditions allow us to solve Eq.~(\ref{eq:vr1})  by Fourier transform (see App.~\ref{sec:solution}).
Next, we summarize the approach and the results.

We begin by recognizing that the total electric current density along $z$ can be written as
\begin{equation}
\label{eq:Jtotal}
J_{\rm tot} = \sum_{n,\chi} j_{n}^\chi + \epsilon \partial_t E_{\rm int},
\end{equation}
where the first term in the r.h.s. stands for the particle current and the second term is the displacement current. 
Out of these parts, the drift current produced by $E_0$ is static (Eq.~(\ref{eq:j_nolight})); the remaining parts are induced by the light pulse and are therefore transients.

 \begin{figure}[t]
  \begin{center}
  % \subcaptionbox{\label{fig:Vt}}[.4\textwidth]
   % \includegraphics[width=0.74\columnwidth]{E_t_png} 
      \includegraphics[width=0.74\columnwidth]{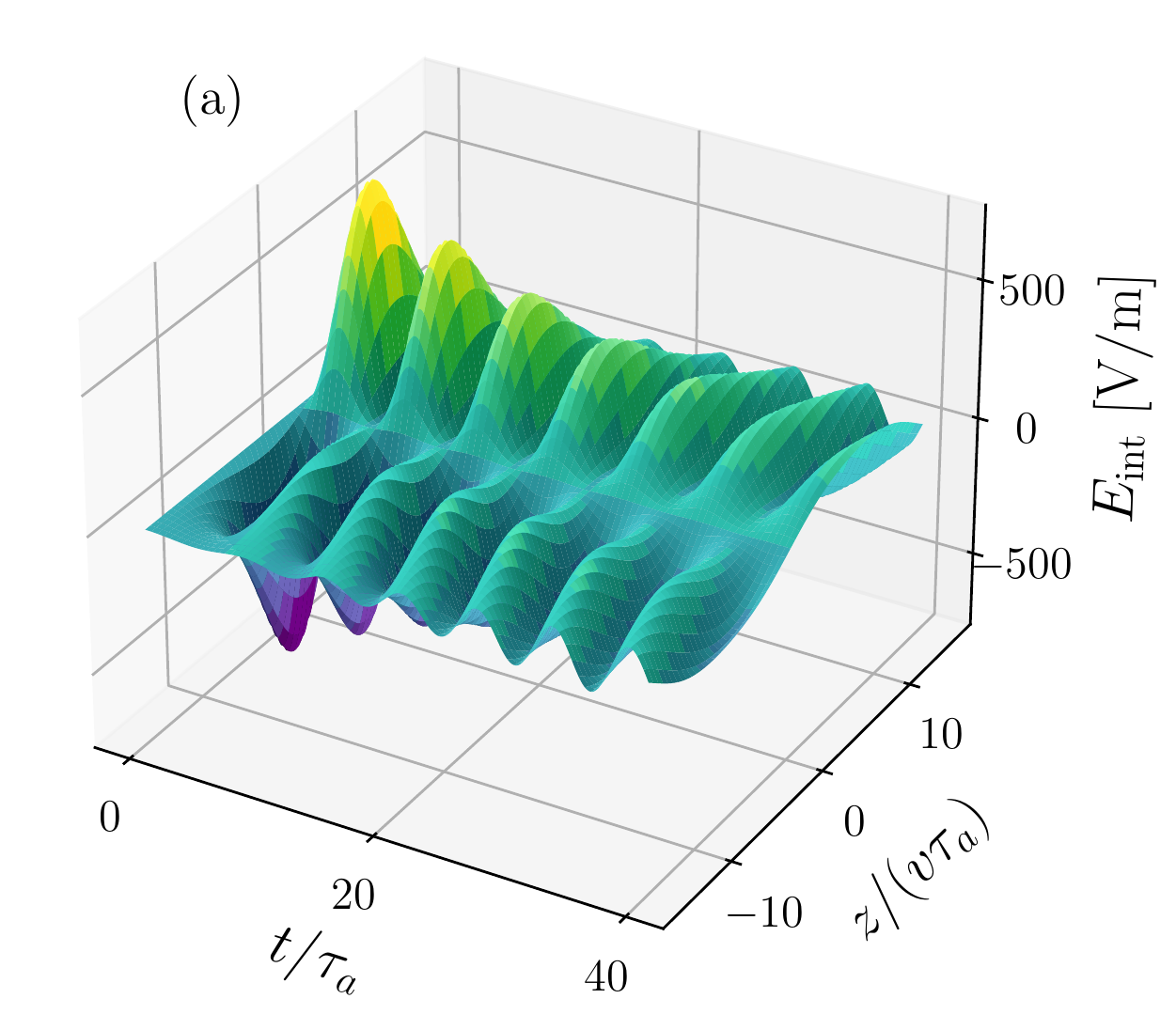} 
      %\subcaptionbox{\label{fig:Et}}[.4\textwidth]
      \includegraphics[width=0.74\columnwidth]{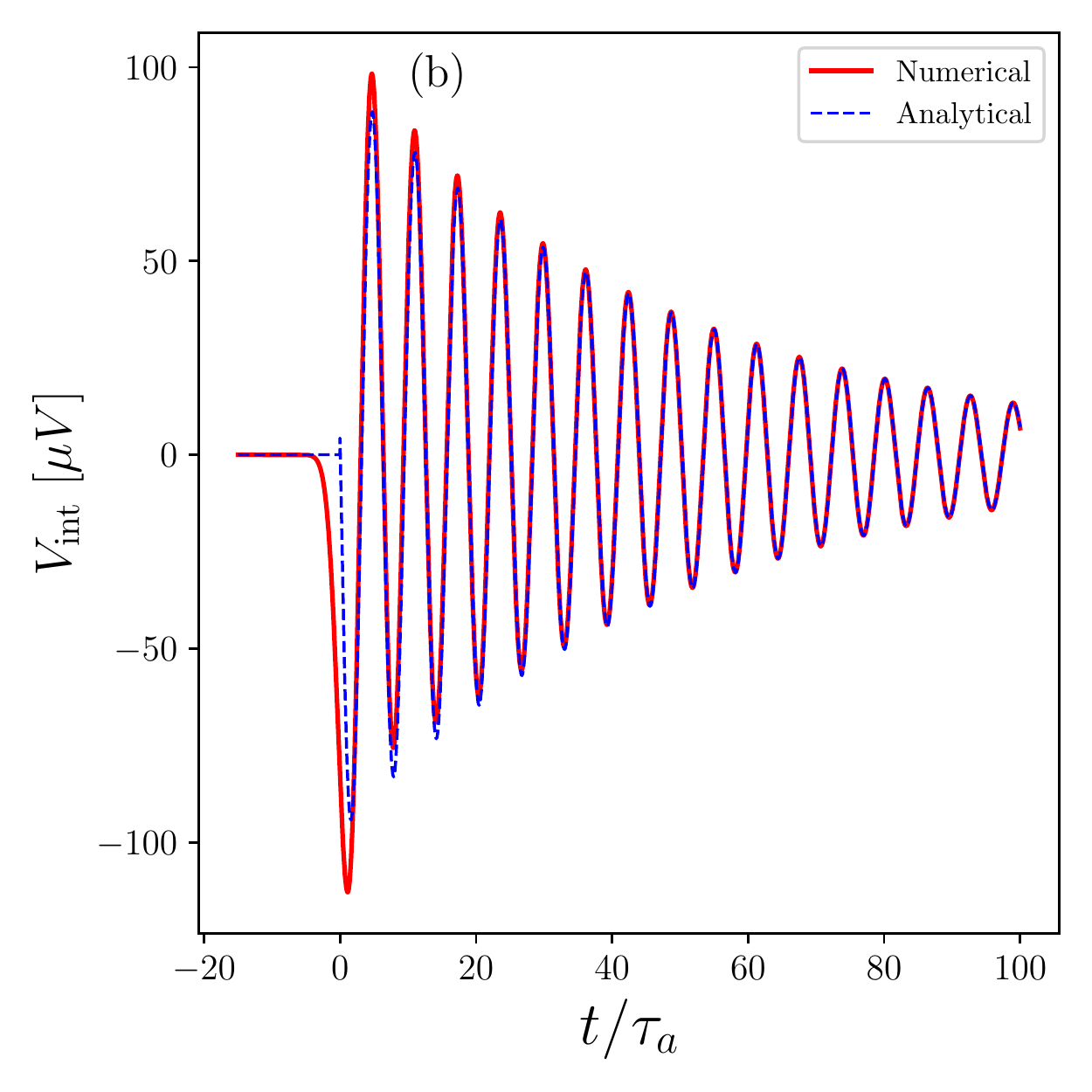}                                                                                                        
\caption{(a) Longitudinal electric field induced by a Gaussian light pulse, centered at position $z=0$ and time $t=0$.
The parameter values (see main text) are: $\tau_a=0.25 {\rm ps}$, $dt=0$, $dz=0.1\mu{\rm m}$, $\tau_r=20 \tau_a$, $\tau_v=50 \tau_a$, $v=10^5 {\rm m/s}$, $\epsilon=30 \epsilon_0$, $v_d=0$, $\overline{g_s} = 10^{14} {\rm m}^{-2}$, $\overline{g_d}=0$.
In this case, $E_{\rm int}(z) = -E_{\rm int}(-z)$.
% is satisfied.
(b) Photovoltage (line-integral of the longitudinal electric field), calculated from Eq.~(\ref{eq:vr_av}). Oscillations at the plasma frequency $\tau_a^{-1}$ stem from the chiral anomaly. The parameter values are the same as in (a), except that $dt=1.5 \tau_a$, $v_d =100 {\rm m/s}$ and $\overline{g_d}=0.2 \overline{g_s}$.
The analytical (Eq.~(\ref{eq:Vsimp2})) and numerical solutions coincide when $\tau_a\ll \tau_r, \tau_v$ and $t\gg dt$.
}
 \label{fig:Vt}
  \end{center}
\end{figure} 

According to Ampère-Maxwell's law, the total electric current density must be divergenceless, which in our model amounts to $\partial_z J_{\rm tot}=0$ \cite{Hawks2015}. Since by construction all transient effects vanish at the contacts, it follows that $J_{\rm tot}$ must be {\em everywhere} static and its value given by Eq.~(\ref{eq:j_nolight}).
In sum, for our boundary conditions, the light pulse has no effect on $J_{\rm tot}$, as though the system were connected to a constant current source.

Contrary to the total current, the voltage drop across the system {\em is} influenced by the light pulse.
In order to see this, we combine Eqs.~(\ref{eq:j1}), (\ref{eq:j0}), (\ref{eq:j_nolight}) and (\ref{eq:Jtotal}) to get 
\begin{equation}
\label{eq:jD}
\epsilon \partial_t E_{\rm int} = -\delta j,
\end{equation}
where
\begin{equation}
\label{eq:delta_j}
\delta j= q v_d \Sigma_1+ q D \partial_z\Sigma_1 + q v \left(\Delta_0 - \sigma_0 E_0/(q v)\right)
\end{equation}
is the particle photocurrent.
In the limit $\tau_a\to 0$, Eq.~(\ref{eq:vr1}) yields $\Sigma_0 = -\Sigma_1$ (charge neutrality for all $z$ and $t$) and hence $E_{\rm int}=0$. By Eq.~(\ref{eq:jD}), $\delta j$ also vanishes when $\tau_a\to 0$. 
This is the ambipolar transport regime \cite{Champlain2011}, in which photoexcited electrons of $n=1$ bands and photoexcited holes of $n=0$ bands track each other locally due to electrical attraction, thereby cancelling their currents.
For $\tau_a\neq 0$, ambipolar transport sets in at $t\gg \tau_a$. 

Integrating Eq.~(\ref{eq:jD}) over the length of the film with the given boundary conditions, we get
\begin{equation}
\label{eq:Vt}
\epsilon  \partial_t V_{\rm int} = \overline{\delta j}= q v_d \overline{\Sigma_1}+ q v \overline{\Delta_0} - \sigma_0 E_0 L,
\end{equation}
where 
$\overline{f}(t) \equiv \int_{-L/2}^{L/2} dz f(z,t)$
and $V_{\rm int}=-\overline{E_{\rm int}}$ is the transient photovoltage.
In Eq.~(\ref{eq:Vt}), $\overline{\Sigma_1}$ and $\overline{\Delta_0}$ obey
\begin{align}
\label{eq:vr_av}
&\left(\partial_t^2 +\frac{2\partial_t}{\tau_v}   +\frac{1}{\tau_a^2}\right) \overline{\Delta_0} 
 =\frac{\sigma_0 E_0 L}{q v \tau_a^2}-\frac{v_d \overline{\Sigma_1}}{v \tau_a^2}+\frac{\partial_t\overline{\Delta_1}}{\tau_r} -\partial_t \overline{g_d}\nonumber\\
& \left(\partial_t +1/\tau_r+2/\tau_v\right)\overline{\Delta_1}= \overline{g_d}\nonumber\\
& \left(\partial_t + 1/\tau_r\right)\overline{\Sigma_1} =\overline{g_s},
\end{align}
which are obtained by a spatial integration of Eq.~(\ref{eq:vr1}).
In the first line of Eq.~(\ref{eq:vr_av}), we have used Eq.~(\ref{eq:Vt}) to replace $\partial_t \overline{E_{\rm int}}$. 

To make further progress, we need additional information about the applied light pulse. As such, we consider $\overline{g^\chi}(t)$ to be Gaussian with a time width $dt$ and an amplitude  $\overline{g_0^\chi}/(\sqrt{2\pi }dt)$, where $\overline{g_0^\chi}$ is a constant.
Then, a simple analytical solution of Eq.~(\ref{eq:vr_av}) is realized in the regime 
$\tau_a \ll (\tau_r,\tau_v)$ 
\footnote{The typical value of $\tau_a$ in the quantum limit is $\sim 0.1 {\rm ps}$, which is orders of magnitude smaller than $\tau_v$ 
\cite{Parameswaran2014, Jadidi2020}. 
For $\tau_r$, we anticipate (see App.~\ref{sec:fermi}) a value $\lesssim 1 {\rm ps}$. Thus, while $\tau_a\ll \tau_r$ is reasonable, $\tau_a\simeq \tau_r$ could also occur. In the latter case, the analytical expression in Eq.~(\ref{eq:Vsimp2}) is less accurate, but remains in semi-quantitative agreement with the numerical solution of Eq.~(\ref{eq:vr_av}).}
and $t\gg dt$  (see App.~\ref{sec:eq18}). 
Substituting that solution in Eq.~(\ref{eq:Vt}), integrating over time and imposing $V_{\rm int}(t\to\infty)= 0$, we get 
%a transient photovoltage
\begin{equation}
\label{eq:Vsimp2}
V_{\rm int}(t) \simeq \frac{q}{\epsilon} \left( v_d\overline{g_{0,s}} - v \overline{g_{0,d}} \right) \tau_a e^{-\frac{dt^2}{2\tau_a^2}-\frac{t}{\tau_v}} \sin\left(t/\tau_a\right) ,
\end{equation}
where 
$\overline{g_{0, s}} = \sum_\chi \overline{g_0^\chi}/2$ and  $\overline{g_{0, d}} = \sum_\chi \chi \overline{g_0^\chi}/2$.
Figure \ref{fig:Vt}b displays Eq.~(\ref{eq:Vsimp2}), together with the numerical solution obtained from Eqs.~(\ref{eq:Vt}) and (\ref{eq:vr_av}).
Figure \ref{fig:Vt}a shows the photoinduced internal electric field, calculated numerically from Eq.~(\ref{eq:vr1}).

\section{Discussion}

\label{sec:disc}

 Eq.~(\ref{eq:Vsimp2}), the key result of this work, can be interpreted as follows.
A light pulse induces a charge separation in the illuminated region and nearby. 
The resulting electric field, $E_{\rm int}$, creates a neutrality-restoring current through the chiral anomaly, but it overshoots and starts plasma oscillations in the $n=0$ bands.
These oscillations are damped by intervalley scattering, which relaxes the current produced by $E_{\rm int}$.
If crystal symmetry is preserved, $E_{\rm int}(z)$ is odd in $z$ and thus  $V_{\rm int}(t)=0$  (see Fig.~\ref{fig:Vt}a and App.~\ref{sec:solution}).
Breaking the $z\to -z$ symmetry, by either $E_0\neq 0$ or by the shape of the light pulse ($\overline{g_{0,d}}\neq 0$), allows for $V_{\rm int}(t)\neq 0$.
If the pulse is slow ($dt\gg \tau_a$), the ambipolar regime sets in while the sample is being illuminated and $V_{\rm int}$ is suppressed.

Although the {\em oscillations} in $V_{\rm int}$ originate from the chiral anomaly, the latter is not required in order to have $V_{\rm int}\neq 0$.  We can ``turn off'' the chiral anomaly in VR equations by taking $\tau_a\to \infty$.
Experimentally, $\tau_a$ can be increased by either reducing $B_0$ or by rotating ${\bf B}_0$ away from $z$.
As soon as $\tau_a>\tau_v$, the dynamics of $\overline{\Delta_0}$ in Eq.~(\ref{eq:vr_av}) becomes overdamped and hence
$V_{\rm int}$ decays monotonically in time (see also  App.~\ref{sec:weak_field}).
%Ref.~[\onlinecite{sm2022}] (Sec. F) corroborates the presence of  a nonoscillatory $V_{\rm int}$ at weak $B_0$.
%magnetic field. 

The appearance of plasma oscillations under light irradiation may seem surprising: bulk plasmons cannot be directly excited by light because they are longitudinal waves, while light waves are transverse (see e.g. Ref.~[\onlinecite{Mattheakis2017}]). Yet, our result is enabled by an indirect mechanism: the asymmetric propagation of electrons and holes excited by light generates a {\em longitudinal} internal electric field, which can drive plasma oscillations.
Furthermore, this mechanism is not unique to topological semimetals: photoinduced plasma oscillations have been observed in semiconductors of trivial band topology \cite{Kersting1997, Kersting1998, Heyman2001}, where they have attracted much interest as a source of THz radiation.
What sets WSM apart is (i) the role of band topology (tunable via ${\bf B}_0$), (ii) the ability to attain the quantum limit with modest $B_0$, and (iii) the relatively long relaxation time $\tau_v$ (due to the relatively large separation in momentum space between counter-moving electrons in the $n=0$ bands).  

Let us estimate the magnitude of $V_{\rm int}$.
For $B_0\simeq 5$T, $|v|\simeq 10^5 {\rm m/s}$ and $\epsilon=30 \epsilon_0$  (where $\epsilon_0$ is the vacuum permittivity), we have $\tau_a\simeq 0.2$ ps.
Taking $\tau\simeq 0.1$ ps, we get $\mu_1\simeq 0.1~$m$^2{\rm/(V s)}$.
Then, $E_0\simeq 1000~{\rm V/m}$ gives  $v_d\simeq 100~{\rm m/s}$.
Using Eq.~(\ref{eq:sigma0}) and $\tau_v\simeq~10$ ps, the dc current density is $\simeq 3\times 10^7$ A/m$^2$.
For a WSM film of cross section $100 \mu{\rm m} \times 1 \mu{\rm m}$, the dc current is modest ($\sim 3\, {\rm m A}$).
An optical generation rate of $10^{25}$ pairs/(s cm$^{3}$) (in App.~\ref{sec:fermi} we estimate that this rate may be attainable with an optical power of $\sim 1$W),  acting during a time $dt\simeq \tau_a$ within a length $dz\simeq 100 \mu {\rm m}$, gives $\overline{g_{0,s}}\simeq  10^{14} {\rm m}^{-2}$. 
If $\overline{g_{0,d}}=0$, we arrive at $V_{\rm int}(t)\lesssim 1 \mu$V.

There are a few strategies to increase $V_{\rm int}$.
First, increasing the power of the optical pulse may be possible, 
%without deleterious Joule heating
though nonlinear effects neglected in our theory might then need consideration. 
Second, the use of a $p-n$ junction with a built-in electric field will render $E_0$ unnecessary, thereby removing the steady state electric current.
Third, in a WSM with multiple pairs of nodes related by time-reversal, each of such pairs will give additive contributions to $V_{\rm int}$.
Fourth, in chiral topological semimetals, $\overline{g_d}\simeq \overline{g_s} $ due to the low crystal symmetry and thus $V_{\rm int}$ could be a factor $|v/v_d|\gg 1 $ larger than in the estimate of the preceding paragraph (see also Fig. \ref{fig:Vt}b, where $\overline{g_d}\neq 0 $).

\section{Conclusions}
\label{sec:conc}

Solving Van Roosbroeck's equations for topologically nontrivial electronic bands, we have predicted photoinduced plasma oscillations in Weyl semimetals and have elucidated their relation to chiral anomaly. 
%It would be interesting to observe such effect experimentally and to assess its possible application for THz emission.
Our findings suggest that it may be interesting to adapt and apply VR equations in order to model a wide variety of topological microelectronic devices.

There are various possible open questions for further work.
First, we assumed that the temperature of the system remains constant and uniform under laser irradiation. Yet, this may not be accurate for high laser intensities. What is the influence of electronic band topology in the dynamics and spatial profile of the temperature? 
%Appropriately adapted VR equations could be used to address this question.

Second, we assumed that the VR equations can be linearized. This requires small departures from equilibrium and therefore limits the scope of our theory. What is the effect of the topological terms in the nonlinear regime of VR equations?

Third, we considered the effect of the chiral anomaly in the VR equations for bulk electrons in a Weyl semimetal. One could study other materials, in which topological quantities should impact the solutions of VR equations. 
%One could also apply VR equations to topological surface states.

%, in which where topological terms in the bulk, while disregarding surface effects. Yet, bulk topological terms are present in other materials as well. The main example is the axion angle, which appears in three dimensional topological insulators and in axion insulators. What is the impact of the axion term in the solution of VR equations? The problem of applying VR equations to topological surface states in a wide variety of devices remains open as well.

Fourth, we considered a simple device geometry with simple boundary conditions. If topological microelectronic devices  become a technological reality, VR equations augmented with topological terms will need to be solved in more realistic settings using appropriately adapted simulation software.

\acknowledgements
This work has been financially supported by the Canada First Research Excellence Fund, the CNRS-Sherbrooke International Research Laboratory on Quantum Frontiers, the Natural Sciences and Engineering Research Council of Canada (Grant No. RGPIN- 2018-05385), and the Fonds de Recherche du Québec Nature et Technologies.
I.G. thanks D. Morris and T. Szkopek for helpful discussions. 
 
\appendix

\begin{widetext}
 
\section{Drift and diffusion currents for the $n=0$ Landau level}
\label{sec:drift_diffusion}

Combining Eqs.~ (2), (3), (4) and (6) of the main text, the total current in the chiral Landau level, $j_0 = j_0^++j_0^-$, can be rewritten as
\begin{equation}
\label{eq:j0_t}
j_0 + \frac{\tau_v}{2} \frac{\partial j_0}{\partial t} = \sigma_0 E +\frac{1}{2} q v^2\tau_v \frac{\partial(\rho_0^+ + \rho_0^-)}{\partial z} - \frac{1}{4}q v \tau_v (g^+-g^-) - q v \frac{\tau_v}{2\tau_r} (\rho_1^+-\rho_1^-).
\end{equation}
%where we assumed $g^+=g^-$ for simplicity.
The first two terms in the right hand side of this equation are the drift and diffusion currents, respectively, with a diffusion constant $D_0 = v^2 \tau_v/2$. Note that $D_0 = \sigma_0/(e^2 \nu(\epsilon_F))$ in consistency with the Einstein relation, where the conductivity $\sigma_0$ is defined in the main text and $\nu(\epsilon_F)$ is the density of states at the Fermi level.
The last two terms in the right hand side of Eq.~(\ref{eq:j0_t}) are currents that emerge due to an unequal light absorption on the two nodes. The left hand side of the equation contains $j_0$ and the time-derivative of $j_0$. This means that, in general, the relation between the current and the carrier density (or the electric field) is nonlocal in time. 
However, if the current varies slowly on the timescale of $\tau_v$, then $j_0\gg \tau_v \partial j_0/\partial t$ and one arrives at the usual drift-diffusion approximation with a local-in-time relation between the current and the carrier densities (or the electric field). 
%In sum,  the drift-diffusion approximation for the $n=1$ is acceptable if the intravalley relaxation time $\tau$ therein is short enough (i.e. $\omega\tau\ll 1$, where $\omega$ is the largest characteristic frequency of the dynamics). 
Since we are interested in the dynamics at timescales that are shorter than $\tau_v$, we do no neglect the $\partial j_0/\partial t$ term in our analysis. 
%For the $n=0$ Landau levels, the drift-diffusion approximation is less accurate, because $\tau_v$ can be long (several orders of magnitude greater than $\tau$). Thus, we will not discard the $\partial j_0/\partial t$ term in our analysis.

\section{Fermi golden rule estimates for the optical generation rate and radiative recombination time}
\label{sec:fermi}

In this section, we provide a numerical estimate for the optical generation rate from the $n=0$ to the $n=1$ Landau level, at fixed chirality $\chi$.
The, we estimate an upper bound for the radiative recombination time (denoted $\tau_r$ is the main text) from the $n=1$ Landau level to the $n=0$ Landau level.

\subsection{ Fermi golden rule estimate for the optical generation rate}

For the purposes of the estimate, let us first consider a monochromatic light of frequency $\omega$, wave vector ${\bf q}$ and polarization vector $\hat{\bf e}$, whose vector potential is given by
\begin{equation}
{\bf A} = \frac{A_0}{2} \hat{\bf e} \left[e^{i ({\bf q}\cdot{\bf r} -\omega t)} + e^{-i ({\bf q}\cdot{\bf r} -\omega t)} \right].
\end{equation}
From Fermi's golden rule \cite{ridley2013quantum}, the generation rate (in units of 1/(volume $\times$ time)) can be written as 
\begin{equation}
\label{eq:golden}
G_0^\chi = \frac{2\pi}{\hbar} \frac{1}{2\pi l_B^2} \frac{1}{L}\sum_k |\langle \psi_{k \chi 1} | H_R | \psi_{k \chi 0}\rangle |^2 (f_{k\chi 0} -f_{k\chi  1}) \delta(E_{k \chi 1}-E_{k \chi 0}-\hbar\omega),
\end{equation}
where $l_B$ is the magnetic length, $L$ is the system length along $z$, $|\psi_{k \chi n}\rangle$ is the electronic eigenstate for the $n-$th Landau level of chirality $\chi$ at wave vector $k$ along $z$, $E_{k n}$ is the corresponding energy, $f_{k \chi n}$ is the Fermi-Dirac occupation factor, 
\begin{equation}
H_R = -\frac{e A_0}{2} e^{i {\bf q}\cdot{\bf r}} \hat{\bf e}\cdot{\bf v}  
\end{equation}
is the light-matter coupling Hamiltonian, and ${\bf v}$ is the electronic velocity operator.
In Eq.~(\ref{eq:golden}), we have neglected the photon wave vector in the electronic interband transitions. 
This is appropriate for the THz frequency light we are interested in (the energy separation between the $n=0$ and $n=1$ for a field of a few Tesla is in the THz regime).

For Weyl fermions, $\langle \psi_{k \chi 1} | {\bf v} | \psi_{k \chi 0}\rangle$ and $E_{k\chi n}$ can be obtained analytically (see e.g. \cite{Parent2020}).
Thereafter, the integration over $k$ in Eq.~(\ref{eq:golden}) can also be carried out analytically, using the Dirac delta function.
Thus, for a polarization vector along $x$, we get
\begin{equation}
G_0^\chi \simeq \frac{e^2 v A_0^2}{16 \pi \hbar^2 l_B^2} 
\end{equation}
where we have used $L^{-1} \sum_k \simeq \int dk/(2 \pi)$ and we have assumed zero temperature.
This expression can be rewritten in terms of the optical power of the laser.
The connection follows from \cite{griffiths2014}
\begin{equation}
\frac{P_{\rm op}}{S} = \frac{ c \epsilon_0 n}{2} E_0^2,
\end{equation}
where $S$ is the area of the illuminated region, $c$ is the speed of light in vacuum, $\epsilon_0$ is the vacuum dielectric constant, $E_0 = \omega A_0$ is the strength of the electric field and $n$ is the refractive index of the WSM. 
Then,
\begin{equation}
\label{eq:G_mono}
G_0^\chi \simeq \frac{e^2 v P_{\rm op}}{8 \pi \epsilon_0 n c  \hbar^2 l_B^2 \omega^2 S}.
\end{equation}

For a given laser power, we can maximize $G_0^\chi$ by making $S$ as small as possible. 
Considering the diffraction limit, the minimum value of $S$ is given by $(\lambda f)^2$, where $\lambda=2\pi c/\omega$ is the wavelength of the light and $f$ is a dimensionless number (the so-called $f$-number of the lens used to focus the light on the sample) \cite{siegman1986}. 
It follows that 
\begin{equation}
G_0^\chi \simeq \frac{e^2 v P_{\rm op}}{32 \pi^3 \epsilon_0 n c^3  \hbar^2 l_B^2 f^2}\simeq 10^{27} \frac{P_{\rm op}[{\rm W}]}{f^2} {\rm cm}^{-3} {\rm s}^{-1},  
\end{equation}
where we have used $v\simeq 10^5 {\rm m/s}$, $n\simeq \sqrt{30}$ and $B=5 {\rm T}$.
Thus, for $f\sim 1-10$, an optical power of $10-1000 \, {\rm mW}$ leads to an optical generation rate of $10^{25} {\rm cm}^{-3} {\rm s}^{-1}$.

The preceding numerical estimate applies for a monochromatic light beam. In reality, since we are interested in a light pulse of duration $dt$ in time, there will be a spread of frequencies of the order of $1/dt$. For each of the frequencies involved in the light beam, we can use the estimate above, with the proviso that $P_{\rm op}$ is the power contained in a specific frequency. The total generation rate is then obtained by integrating the rate over all relevant frequencies. 
We define a density of optical generation at frequency $\Omega$,
\begin{equation}
\label{eq:G_script}
\mathcal{G}_0^\chi(\Omega) \simeq \frac{e^2 v \mathcal{P}_{\rm op}(\Omega)}{8 \pi \epsilon_0 n c  \hbar^2 l_B^2 \Omega^2 S},
\end{equation}
such that 
\begin{equation}
\label{eq:G_int}
G_0^\chi = \int_{\rm gap}^\infty d\Omega \mathcal{G}_0^\chi(\Omega).
\end{equation}
In Eq.~(\ref{eq:G_script}),  $\mathcal{P}_{\rm op}$ is the power density, such that $\mathcal{P}_{\rm op}(\Omega) d\Omega$ describes the power in the frequency interval $(\Omega, \Omega+d\Omega)$.
In Eq.~(\ref{eq:G_int}), the lower bound of the integral is equal to the optical gap (the minimum photon frequency that can induce a vertical interband transition between the $n=0$ and $n=1$ Landau levels).
We recover the result (\ref{eq:G_mono}) for a monochromatic light when $\mathcal{P}_{\rm op}(\Omega) = P_{\rm op} \delta(\Omega-\omega)$.
We can generalize this density to the case of a light pulse of duration $dt$ via
\begin{equation}
\mathcal{P}_{\rm op}(\Omega) = \frac{2 dt}{\sqrt{\pi} \left(1+ {\rm erf}(\omega dt)\right)} P_{\rm op} e^{-(\Omega-\omega)^2 dt^2},
\end{equation}
normalized such that $\int_0^\infty \mathcal{P}_{\rm op}(\Omega) d\Omega = P_{\rm op}$.
For an infinitely long pulse, we recover the monochromatic light beam.
Then, if we still assume that $S\simeq (2\pi c f/\Omega)^2$ and if we neglect the frequency-dependence of the refractive index within a frequency range of order $1/dt$, we obtain the following generalization of Eq.~(\ref{eq:G_mono}):
\begin{align}
G_0^\chi &\simeq \frac{e^2 v P_{\rm op}}{32 \pi^3 \epsilon_0 n c^3  \hbar^2 l_B^2 f^2} \frac{ 2 dt}{\sqrt{\pi} \left(1+ {\rm erf}(\omega dt)\right)} \int_{\rm gap}^{\infty} e^{-(\Omega-\omega)^2 dt^2} d\Omega\nonumber\\
&=\frac{e^2 v P_{\rm op}}{32 \pi^3 \epsilon_0 n c^3  \hbar^2 l_B^2 f^2} \frac{1+{\rm erf}[(\omega-{\rm gap}) dt]}{1+ {\rm erf}(\omega dt)}
\end{align}
Thus, the result of the monochromatic case is corrected by a factor 
\begin{equation}
F = \frac{1+{\rm erf}[(\omega-{\rm gap}) dt]}{1+ {\rm erf}(\omega dt)}.
\end{equation}
This factor is smaller than unity (see Fig. \ref{fig:erf}), with $F\simeq 0$ when $\omega\ll 1/dt \ll {\rm gap}$ and $F\to 1$ when $\omega\gg {\rm gap}$.
In our case of interest, where $1/dt \simeq {\rm gap}\simeq \omega$, we have $F\simeq 0.5$. Thus, the estimate from the monochromatic case is not qualitatively changed.

\begin{figure}[t]
  \begin{center}
    \includegraphics[width=0.8\columnwidth]{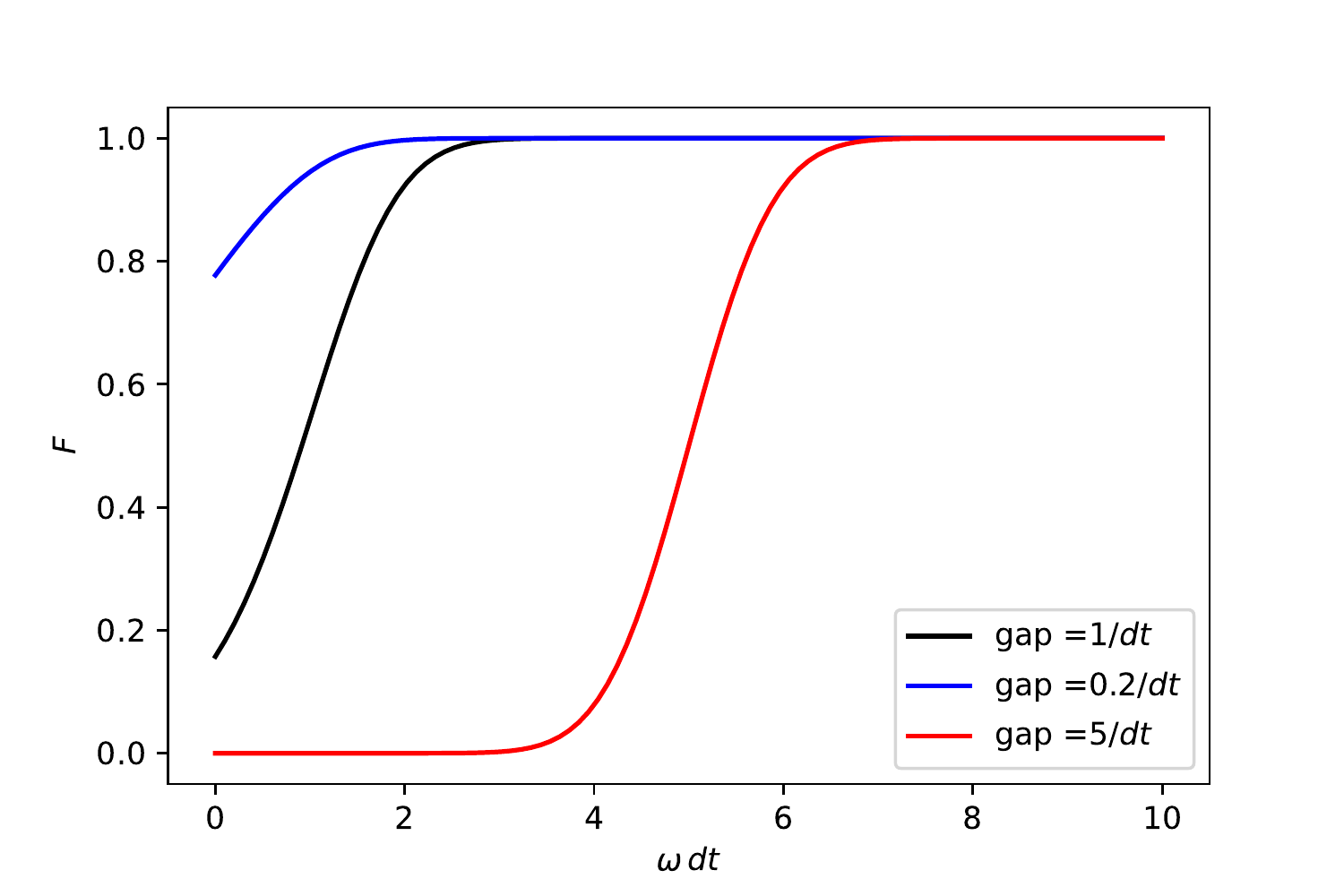}                                                                                               
\caption{}
 \label{fig:erf}
  \end{center}
\end{figure} 

\subsection{Fermi golden rule estimate for the radiative recombination time}

%The preceding estimate for the optical generation rate is helpful to attain an estimate of the radiative recombination rate. 
The Fermi golden rule expression for the rate of spontaneous radiative electron-hole recombination between $n=0$ and $n=1$ Landau levels is similar to Eq.~(\ref{eq:golden}), except that we need to divide it by the number of photons impinging on the sample.
Using the relation between $A_0^2$ and the photon number \cite{ridley2013quantum}, we have
\begin{equation}
\label{eq:recorate}
\frac{1}{\tau_r} > \frac{e^2 v \hbar}{8 \pi^2 \hbar^2 l_B^2 \epsilon \omega}\simeq 10^{12} {\rm s}^{-1},
\end{equation}
where we have assumed $\omega\simeq v/l_B$ (typical frequency for interband transition between $n=0$ and $n=1$), $B=5 {\rm T}$, $v=10^5 {\rm m/s}$, and $\epsilon=30 \epsilon_0$. 
This estimate is a lower bound for $1/\tau_r$, as it assumes a fixed polarization of the emitted photon (the average over all polarizations will result in a multiplication by a factor of order one) and it neglects the phonon-mediated recombination.
Thus, $\tau_r\lesssim 10^{-12} {\rm s}$.

\section{Linearized van Roosbroeck equations}
\label{sec:linearized}

In this Appendix, we show the details leading to Eq. (11)  in the main text.
First, for the $n=1$ bands, we take a sum of the $\chi=1$ and $\chi=-1$ equations in order to arrive at
\begin{equation}
\label{eq:S1}
\mu_1 \left(\Sigma_1 +\rho_{1,\rm eq}^+ + \rho_{1,\rm eq}^- \right) \partial_z E + \mu_1 E \partial_z\Sigma_1 + D_1 \partial^2_z\Sigma_1 - \partial_t\Sigma_1 = \frac{\Sigma_1}{\tau_r} -\frac{1}{2} (g^+ +g^-),
\end{equation}
 where $E$ is the $z-$component of the total electric field and $\Sigma_1 \equiv \rho_1^+ + \rho_1^- -\rho_{1,\rm eq}^+ - \rho_{1,\rm eq}^-$.
 %, $\Sigma_z^1 \equiv \partial_z \Sigma^1$, $\Sigma^1_{zz} \equiv \partial^2_z \Sigma^1$ and $E_z\equiv \partial_z E$ (do not confuse $E_z$ with the $z$ component of the electric field; the $z$ component is already denoted as $E$).
 Equation~(\ref{eq:S1}) is a nonlinear differential equation. In order to make analytical progress, we will linearize by assuming that the departure of the carrier distribution from equilibrium is not strong. For instance, assuming that both $\rho_{1,\rm eq}^\chi$ and $\Sigma_1$ are small (this is justified in the quantum limit at low temperature, provided that the effect of the light in the carrier distribution is not strong), we will neglect the first term in Eq.~(\ref{eq:S1}).
Similarly,  we approximate  $\mu_1 E \partial_z \Sigma_1 \simeq  \mu_1 E_0 \partial_z \Sigma_1$.
Consequently,  the linearized Eq.~(\ref{eq:S1}) reads
\begin{equation}
\label{eq:S1b}
\mu_1 E_0 \partial_z\Sigma_1 + D_1 \partial^2_z\Sigma_1 - \partial_t\Sigma_1 \simeq \frac{\Sigma_1}{\tau_r} -\frac{1}{2} (g^+ +g^-).
\end{equation}

Second, still for the $n=1$ bands, we take the difference of the $\chi=1$ and $\chi=-1$ equations in order to arrive at
\begin{equation}
\label{eq:D1}
\mu_1 E_0 \partial_z \Delta_1 + D_1 \partial^2_z\Delta_1 - \partial_t\Delta_1 \simeq \frac{\Delta_1}{\tau_r} +\frac{2 \Delta_1}{\tau_v} -\frac{1}{2} (g^+ -g^-),
\end{equation}
where $\Delta_1\equiv \rho_1^+ - \rho_1^- -\rho_{1,\rm eq}^+ + \rho_{1,\rm eq}^-$ and we have already performed the linearization. 
Note that Eq.~(\ref{eq:D1}) is the same as Eq.~(\ref{eq:S1b}), except for the source term and the relaxation term: only $g^+\neq g^-$ can lead to a nonzero $\Delta_1$ and, unlike in the case of $\Sigma_1$ (as the two nodes are mirror-partners in equilibrium), elastic intervalley scattering relaxes $\Delta_1$.
While $\Sigma_1$ is driven by $g^+ + g^-$, $\Delta_1$ is driven by $g^+-g^-$. 
Thus, in the absence of $E_0$, we have no source term for $\Delta_1$ and thus we will have $\Delta_1=0$ for all times and positions.

Third, we discuss the equations for $n=0$. On one hand, Eq.~(3) of the main text leads to
\begin{align}
\label{eq:1}
v\partial_z \rho_0^+ -\partial_t\rho_0^+ &= \frac{\rho_0^+-\rho_0^-}{\tau_v} - \frac{\rho_1^+-\rho_{1,\rm eq}^+}{\tau_r} - \frac{q^2}{4\pi^2\hbar^2} E B_0 +\frac{g^+}{2}\\
\label{eq:2}
-v\partial_z \rho_0^- -\partial_t\rho_0^- &= \frac{\rho_0^--\rho_0^+}{\tau_v} - \frac{\rho_1^--\rho_{1,\rm eq}^-}{\tau_r} + \frac{q^2}{4\pi^2\hbar^2} E B_0 +\frac{g^-}{2}.
\end{align}
Summing Eqs.~(\ref{eq:1}) and (\ref{eq:2}) yields
\begin{equation}
\label{eq:3}
v\partial_z \Delta_0- \partial_t\Sigma_0= -\frac{\Sigma_1}{\tau_r} + g_s,
\end{equation}
while taking the difference between Eqs.~(\ref{eq:1}) and (\ref{eq:2}) gives
\begin{equation}
\label{eq:4}
v\partial_z\Sigma_0 - \partial_t\Delta_0 = 2\frac{\Delta_0}{\tau_v} - 2\frac{q^2}{4\pi^2\hbar^2} E B_0 -\frac{\Delta_1}{\tau_r}+ g_d.
\end{equation}
%where we have used $\rho_{1,\rm eq}^+ = \rho_{1,\rm eq}^-$.

\section{Solution of the van Roosbroeck equations in Fourier space}
\label{sec:solution}

The simplest situation in which Eq.~(11) in the main text can be solved consists of an infinitely long system such that, at $z\to\pm \infty$ or $t\to\pm \infty$, the influence of the light pulse on the carrier densities is negligible. 
Then, we may define the Fourier transform of a function $f(z,t)$ as
\begin{equation}
\tilde{f}(k,\omega)=\int_{-\infty}^\infty dz e^{i k z} \int_{-\infty}^{\infty} dt  e^{-i\omega t} f(z,t),
\end{equation}
and use relations such as
\begin{align}
&\int dt dz  e^{i k z- i \omega t} \partial_z f(z,t) =  \int dt dz  \partial_z \left(e^{i k z- i \omega t} f(z,t)\right) -  i k \int dt dz  e^{i k z- i \omega t} f(z,t) =  - i k \tilde{f}(k,\omega)\\
&\int dt dz  e^{i k z- i \omega t} \partial_t f(z,t) =  \int dt dz  \partial_t \left(e^{i k z- i \omega t} f(z,t)\right) +  i \omega \int dt dz  e^{i k z- i \omega t} f(z,t) =  i\omega \tilde{f}(k,\omega).
\end{align}
%where we have used the boundary condition $\Sigma^1(z\to\pm\infty,t) = 0$ and we have defined $\tilde\Sigma^1(k,\omega) =\int dt dz  e^{i k z- i \omega t} \Sigma^1 (z,t)$.
Proceeding in this way, Eq.~(11) of the main text can be rewritten as
\begin{align}
\label{eq:vr_ft}
& -i k \tilde{E}_{\rm int} = - \frac{q}{\epsilon} (\tilde{\Sigma}_0 + \tilde{\Sigma}_1)\nonumber\\
& \left(-i k v_d -D_1 k^2 - i\omega-\frac{1}{\tau_r}\right) \tilde{\Sigma}_1 = -\tilde{g_s}\nonumber\\%frac{1}{2}\left(\tilde{g}^+ + \tilde{g}^-\right)\nonumber\\
& \left(-i k v_d -D_1 k^2 - i\omega-\frac{1}{\tau_r}-\frac{2}{\tau_v}\right) \tilde{\Delta}_1 = -\tilde{g}_d\nonumber\\
&-i v k \tilde{\Delta}_0 - i\omega\tilde{\Sigma}_0 = -\frac{\tilde{\Sigma}_1}{\tau_r} + \tilde{g}_s\nonumber\\
&- i v k \tilde{\Sigma}_0 - \left(i\omega+\frac{2}{\tau_v}\right)\tilde{\Delta}_0 = \frac{\epsilon\tilde{E}}{q v \tau_a^2} - \frac{\tilde{\Delta}_1}{\tau_r} + \tilde{g}_d,
%&\left(-v^2 k^2 +\omega^2 -\frac{2 i \omega}{\tau_v} - \frac{1}{\tau_a^2}\right) \tilde{\Sigma}^0 = \left(-\frac{i \omega}{\tau_r} -\frac{2}{\tau_r\tau_v} + \frac{1}{\tau_a^2}\right)\tilde{\Sigma}^1 
%+\frac{i v k}{\tau_r} \tilde{\Delta}^1+\left(\frac{i\omega}{2} + \frac{1}{\tau_v}\right) \left(\tilde{g}^+ + \tilde{g}^-\right) -\frac{i v k}{2}\left(\tilde{g}^+-\tilde{g}^-\right)\nonumber\\
%&\left(-v^2 k^2 +\omega^2 -\frac{2 i\omega}{\tau_v}\right) \tilde{\Delta}^0 = \frac{ i v k}{\tau_r} \tilde{\Sigma}^1 - \frac{i \omega}{\tau_r} \tilde{\Delta}^1-\frac{i\omega}{v q \tau_a^2} \epsilon (\tilde{E}_{\rm int}+\tilde{E}_{\rm light}) -\frac{ i v k}{2} \left(\tilde{g}^+ + \tilde{g}^-\right) +\frac{i\omega}{2}\left(\tilde{g}^+-\tilde{g}^-\right).
\end{align}
where $\tilde{E}= \tilde{E}_{\rm light} + \tilde{E}_{\rm int}$. These are now algebraic equations whose solution is straightforward (though cumbersome).
%{\bf Remark for later reference: When $g^+\neq g^-$, the displacement current can still be written as the difference between total current and particle current, but one must also include a "photogalvanic current" term that is known to arise in noncentrosymmetric systems (which would be our scenario if $g^+\neq g^-$).}

For simplicity, let us first neglect the difference between $g^+$ and $g^-$ (i.e. assume $g^+ \simeq g^- \equiv g$).
This immediately implies $\Delta_1=0$, and
\begin{align}
\label{eq:ft}
\tilde{\Sigma}_1 &= \frac{\tilde{g} \tau_r}{1+D_1 k^2 \tau_r + i k v_d \tau_r + i \omega\tau_r}\nonumber\\
\tilde{\Sigma}_0 &= \frac{\tilde{g} \tau_r \left(-\tau_v/\tau_a^2-2 i \omega+\omega^2\tau_v + D_1 k^2 (-2-i\omega\tau_v) + v_d k(-2 i+\omega\tau_v)\right)}{\left(1+D_1 k^2 \tau_r + i k v_d \tau_r + i \omega\tau_r\right)\left(\tau_v/\tau_a^2 + k^2v^2\tau_v +2 i \omega - \omega^2\tau_v\right)}\nonumber\\
\tilde{\Delta}_0&=\frac{\tilde{g}\tau_r\tau_v \left(i(1/\tau_a^2+v^2k^2)(D_1k+i v_d)-v^2 k \omega\right)}{v\left(1+D_1 k^2 \tau_r + i k v_d \tau_r + i \omega\tau_r\right)\left(\tau_v/\tau_a^2 + k^2v^2\tau_v +2 i \omega - \omega^2\tau_v\right)}+ i \frac{\epsilon \tilde{E}_{\rm light} \omega}{q v \tau_a^2\left(k^2 v^2 -\omega^2 + 2 i \omega/\tau_v\right)}\nonumber\\
\tilde{E}_{\rm int} &=-\frac{q}{\epsilon}\frac{\tilde{g}\tau_r\left(2 v_d+i \tau_v(k v^2+\omega v_d)+ D_1 k (-2 i+\omega\tau_v)\right)}{\left(1+D_1 k^2 \tau_r+ i \tau_r(v_d k+\omega)\right)\left(\tau_v/\tau_a^2+ v^2 k^2 \tau_v +\omega(2 i-\omega\tau_v)\right)}.
\end{align}
The last term in the third line of Eq.~(\ref{eq:ft}) comes from the chiral anomaly induced by the electric field of light within the chiral Landau levels. 
This term would be present even in the absence of light-induced interband transitions. 
In particular, it does not involve $\tau_r$, because it is independent of the interband absorption rate.
Note that $\tilde{E}_{\rm light}$ is the Fourier transform of the $z-$component of $E_{\rm light}$. 
If the electric field of the incident light is oriented perpendicular to the magnetic field, the last term of the third line of Eq.~(\ref{eq:ft}) will be absent.
This is the situation we will adopt from now on, in order to distinguish the previously known physics from our new predictions.
%{\bf I am tempted to restrict the theory to the latter configuration. The message would be that even in the absence of the ordinary chiral anomaly induced by the electric field of light through intraband chiral pumping, we still have chiral anomaly effects coming from interband excitation of carriers. Note: strictly speaking, the polarization of a Gaussian beam is nonuniform. Thus, it may not be exact that the electric field of the light will be perpendicular to $z$ everywhere inside the beam. }

An inverse Fourier transform of Eq.~(\ref{eq:ft}),
\begin{equation}
\label{eq:ift}
f(z,t) =\int_{-\infty}^\infty \frac{dk}{2\pi} e^{-i k z} \int_{-\infty}^{\infty} \frac{d\omega}{2\pi} e^{i\omega t} \tilde{f}(k,\omega),
\end{equation}
 allows us to calculate the time- and space-dependence of $\Sigma_n$, $\Delta_n$ and $E_{\rm int}$. 

For general light pulses, Eq.~(\ref{eq:ift}) must be computed numerically.
%Below, we present analytical approximations valid for certain physically relevant regimes.
Yet, even without calculation it is obvious that $\Delta_0$ is generally nonzero (i.e., there is a valley polarization) despite the fact that light is absorbed with equal intensity in the two valleys. 
The explanation for this peculiarity resides in the fact that chiral Landau levels are unidirectional, with opposite group velocities for opposite chiralities.
Thus, upon light irradiation, holes in the two chiral Landau levels counter propagate, which locally (at each $z$) gives rise to a nonzero $\Delta_0$.

Another useful result can be extracted from Eq.~(\ref{eq:ft}) without any calculation, simply by observing the $k-$dependence of different terms and combining this with Eq.~(\ref{eq:ift}). 
Let us neglect the term proportional to $E_{\rm light}$. Then, if $E_0=0$ and $g(z)=g(-z)$, we have $\Sigma_0(z)=\Sigma_0(-z)$, $\Delta_0(z)=-\Delta_0(-z)$ and $E_{\rm int}(z)=-E_{\rm int}(-z)$. 
This result can be understood from the facts that (i) when $E_0=0$ and $g(z)=g(-z)$, the system has inversion symmetry along $z$; (ii)
the photoexcited holes in chiral Landau levels counter propagate, with holes of opposite chirality going opposite ways.
As a result, there is an excess of holes of positive chirality on $z<0$ and an equal excess of holes of negative chirality on $z>0$, thereby giving rise to  $\Delta_0(z) = -\Delta_0(-z)$.
When $E_0\neq 0$ or $g(z)\neq g(-z)$, inversion symmetry along $z$ is broken, so that $\Sigma_0(z)\neq \Sigma_0(-z)$, $\Delta_0(z)\neq \Delta_0(-z)$ and $E_{\rm int}(z) \neq -E_{\rm int}(-z)$.
The latter asymmetry in the internal electric field is crucial for the development of the transient photovoltage discussed below and also in the main text. 
%The term proportional to $E_{\rm light}$ in Eq.~(\ref{eq:ft}) describes internodal charge pumping solely via the chiral Landau levels.
%On its own, this term would lead to $\Delta_0(z)=\pm \Delta_0(-z)$ if $E_{\rm light}(z)=\pm E_{\rm light}(-z)$.

In order to gain further analytical understanding of Eq.~(\ref{eq:ft}), it is useful to consider some simple limits.
First, we find
\begin{equation}
\label{eq:tauAa}
\lim_{\tau_a\to 0} \left(\tilde{\Sigma}_0+\tilde{\Sigma_1}\right) =0.
\end{equation}
This means that, at timescales that are very long compared to the dielectric relaxation time, the charge neutrality (which was initially perturbed by the fact that the photoexcited holes in the chiral Landau levels and the photoexcited electrons in the nonchiral Landau level propagate at different velocities) will be locally restored. 
In other words, the photoinduced holes in the chiral Landau levels and the photoinduced electrons in the nonchiral Landau level will propagate in lockstep.
In the semiconductor literature, this is known as the "ambipolar transport regime"\cite{Champlain2011}.
We note that the chiral anomaly term in the continuity equations is crucial in order to reach the ambipolar transport regime.
Likewise, if we disregard the term proportional to $E_{\rm light}$, we find
\begin{equation}
\label{eq:tauAb}
\lim_{\tau_a\to 0} \tilde{\Delta}_0 = \tilde{g} \frac{(i D_1 k - v_d)\tau_r}{v (1+D_1 k^2\tau_r + i \tau_r(v_d k+\omega))}.
\end{equation}
This result is consistent with the ambipolar transport regime: it leads to the fact that the current due to excess holes in the chiral Landau levels ($q v \Delta_0$) exactly cancels with the current from excess electrons in the nonchiral Landau level ($q v_d \Sigma_0 + q D_1 \partial_z \Sigma_1$); the cancellation can be verified directly in Fourier space $(k,\omega)$.
Concerning the term proportional to $E_{\rm light}$ in the last line of Eq.~(\ref{eq:ft}), it describes an oscillatory current of intraband origin; there is no spatial charge separation (or internal electric field) associated to it, and thus is not relevant for the emergence of the ambipolar regime.

Second, a concrete situation of interest is that of a delta function pulse in space and time, i.e.  $g=g_0 \delta(z)\delta(t)$ for a constant $g_0$, which results in  $\tilde{g}=g_0$ independent $k$ and $\omega$.
Then, using the residue theorem,
\begin{equation}
\Sigma_1(z,t) =\int_{-\infty}^\infty \frac{dk}{2\pi} e^{-i k z} \int_{-\infty}^{\infty} \frac{d\omega}{2\pi} e^{i\omega t} \tilde{\Sigma}_1(k,\omega)=\frac{g_0}{2\sqrt{\pi D_1 t}} e^{-(z+v_d t)^2/(4 D_1 t)} e^{-t/\tau_r},
\end{equation}
which agrees with the results in standard semiconductor textbooks \cite{McKelvey1984}.
The inverse Fourier transforms for $\tilde{\Sigma}_0$ and $\tilde{\Delta}_0$ are analytically more cumbersome. 
In order to do some reality checks, we consider the limit $(\tau_a,\tau_r,\tau_v) \to \infty$.
Then, we get
\begin{align}
\label{eq:norel}
%\tilde{\Sigma}^1 &\simeq \frac{\tilde{g}}{D_1 k^2 + i k v_d  + i \omega}\nonumber\\
\tilde{\Sigma}_0 &\simeq \frac{i\tilde{g}\omega}{\omega^2-v^2 k^2-i 0^+\rm{sign}(\omega)}\nonumber\\
\tilde{\Delta}_0 &\simeq-\frac{i \tilde{g} v k} {\omega^2-v^2 k^2-i 0^+\rm{sign}(\omega)},
\end{align}
where $0^+=1/\tau_v$ is an infinitesimal positive number (kept to ensure the causality of the solution).
Recalling that $\tilde{g}\propto B$, the expressions for $\tilde{\Sigma}^0$ and $\tilde{\Delta}^0$ are proportional, respectively, to the scalar and axial density response functions in the quantum limit \cite{Rinkel2019}. Because the only spatial variations in our problem take place along the direction of the magnetic field ($z$), the relevant response functions are those with zero transverse wave vector.
%{\bf Mention that $g_0$ is proportional to the Landau level degeneracy}
The axial response function appearing in the expression for $\tilde{\Delta}^0$ is associated to the chiral anomaly \cite{Rinkel2019}, and its form remains unchanged at weaker magnetic fields when the system is no longer in the quantum limit.

Now let us compute the inverse Fourier transform of Eq.~(\ref{eq:norel}). For delta function pulses, we obtain
\begin{align}
\label{eq:norel2}
%\tilde{\Sigma}^1 &\simeq \frac{\tilde{g}}{D_1 k^2 + i k v_d  + i \omega}\nonumber\\
\lim_{(\tau_a,\tau_r,\tau_v) \to \infty} \Sigma_0(z,t) &\simeq -\frac{g_0}{2} \left[\delta(z+v t) + \delta(z- v t)\right]\nonumber\\
\lim_{(\tau_a,\tau_r,\tau_v) \to \infty} \Delta_0 (z,t) &\simeq -\frac{g_0}{2} \left[\delta(z+v t) - \delta(z- v t)\right].
\end{align} 
For finite $(\tau_a,\tau_r,\tau_v)$, this result remains relevant at timescales that are short compared to $\tau_a$, $\tau_r$ and $\tau_v$.
In that regime, the excess holes induced optically in the chiral Landau levels counter propagate without attenuation, with opposite group velocities for carriers of opposite chirality. 
The overall negative sign in the first line of Eq.~(\ref{eq:norel2}) is due to the fact that the optical pulse removes electrons from the chiral Landau level in order to put them in the nonchiral Landau levels.
This explains the two lines of Eq.~(\ref{eq:norel2}).
 It is remarkable that, in this regime, the dynamics of carriers in the chiral Landau levels is decoupled from the dynamics of charge carriers in the nonchiral Landau levels.
For timescales long compared to $\tau_a$, one no longer has two decoupled and counter propagating delta functions. Instead, as shown by Eqs. (\ref{eq:tauAa}) and (\ref{eq:tauAb}), the excess charges in the chiral Landau levels trail the excess charges in the nonchiral Landau level, in order to realize the ambipolar regime (local charge neutrality and zero net current due to photoinduced excess charges).  
%{\bf Points of discussion to be added: axial and polar response functions, counterpropagating delta functions in the chiral Landau levels.}

\section{Details on the derivation of Eq. (18) in the main text}
\label{sec:eq18}

For simplicity, we begin by assuming a delta-function light pulse in time, i.e. $\overline{g^\pm} (t)= \overline{g_0^\pm} \delta(t)$, with a constant $\overline{g_0^\pm}$ (the "bar" notation has been introduced in the main text).
Then, the solution of Eq.~(17) in the main text with the appropriate boundary conditions is obtained by using the residue theorem, 
\begin{align}
\label{eq:solf}
\overline{\Sigma_1} &=\Theta(t)\frac{\overline{g_0^+} + \overline{g_0^-}}{2} e^{-t/\tau_r}\nonumber\\
\overline{\delta\Delta_0} &=\Theta(t)\frac{(\overline{g_0^+} + \overline{g_0^-}) v_d}{2 v} \frac{\tau_d^2}{\tau_a^2} \frac{1}{(1+\Omega^2\tau_d^2)} \left(-e^{-t/\tau_r}+e^{-t/\tau_v}\left(\cos(\Omega t) +\frac{\sin(\Omega t)}{\Omega\tau_d} \right)\right)\nonumber\\
&+\Theta(t) \frac{(\overline{g_0^+} - \overline{g_0^-})}{2}\frac{\tau_s^2}{\tau_a^2}\frac{1}{(1+\Omega^2\tau_s^2)} \left(-\frac{\tau_a^2}{\tau_r}\left(\frac{1}{\tau_r}+\frac{2}{\tau_v}\right) e^{-t(1/\tau_r+2/\tau_v)}+e^{-t/\tau_v} \left(-\cos(\Omega t)
+\frac{\sin(\Omega t)}{\Omega\tau_s}\right)\right),
\end{align}
where $\Theta(t)$ is the Heaviside function, $\overline{\delta\Delta_0} \equiv \overline{\Delta_0} -\sigma_0 E_0 L/(q v)$ and
\begin{align}
\Omega &\equiv \sqrt{\frac{1}{\tau_a^2} - \frac{1}{\tau_v^2}}\nonumber\\
\frac{1}{\tau_{d (s)}} &\equiv \frac{1}{\tau_v} - (+)\frac{1}{\tau_r}.
\end{align}
It is worth noting that $\Omega$ is the plasmon frequency in the quantum limit of a Weyl semimetal, renormalized by the damping $1/\tau_v$ \cite{Parent2020}.
Thus, we learn that $\overline{\Delta_0}$ oscillates at the plasma frequency.

In the regime in which $\tau_a\ll (\tau_r,\tau_v)$, we obtain the following approximate expression for Eq.~(16) of the main text:
\begin{align}
\label{eq:dVdt0}
\epsilon\partial_t  V_{\rm int}&\simeq \Theta(t) q v_d\frac{(\overline{g_0^+} + \overline{g_0^-})}{2}  e^{-t/\tau_v}\left(\cos(\Omega t) +\frac{\sin(\Omega t)}{\Omega\tau_d} \right)\nonumber\\
&+\Theta(t) q v \frac{(\overline{g_0^+} - \overline{g_0^-})}{2} e^{-t/\tau_v} \left(-\cos(\Omega t)+\frac{\sin(\Omega t)}{\Omega \tau_s}\right)\nonumber\\
&+\Theta(t)q v_d \frac{(\overline{g_0^+} + \overline{g_0^-})}{2} \left(-\frac{2}{\tau_v}+\frac{1}{\tau_r}\right)\frac{\tau_a^2}{\tau_r}e^{-t/\tau_r}\nonumber\\
&-\Theta(t) q v \frac{(\overline{g_0^+} - \overline{g_0^-})}{2}\left(\frac{2}{\tau_v}+\frac{1}{\tau_r}\right)\frac{\tau_a^2}{\tau_r}e^{-t(1/\tau_r+2/\tau_v)}.
% +  \Theta(t) q v_d \frac{g^+ + g^-}{2}\frac{1}{1+\Omega^2\tau_d^2}\left(1-\frac{\tau_d^2}{\tau_v^2}\right) e^{-t/\tau_r}.
\end{align}
%We see that, as $\tau_a\to 0$, $\partial_t V_{\rm int}$ oscillates very rapidly and thus averages out, as expected in the ambipolar regime.
Integrating this over time, we have 
\begin{equation}
V_{\rm int}(t) \simeq \left\{\begin{array}{cc} V_{\rm int}(\infty) + A e^{-t/\tau_v}\sin(t/\tau_a)+ B e^{-t/\tau_v} \cos(t/\tau_a)+ C e^{-t/\tau_r}+D e^{-t/\tau_r-2 t/\tau_v}, &\text{ for } t>0\\
V_{\rm int}(0^-), &\text{ for } t<0,
\end{array}\right.
\end{equation}
where $V_{\rm int}(\infty)$ and $V_{\rm int}(0^-)$ are integration constants (corresponding to the values of $V_{\rm int}$ at $t\to\infty$ and $t\to 0^-$, respectively), and
\begin{align}
A &= \frac{q v_d}{\epsilon} \frac{\overline{g_0^+} + \overline{g_0^-}}{2} \tau_a-\frac{q v}{\epsilon} \frac{\overline{g_0^+} - \overline{g_0^-}}{2} \tau_a\nonumber\\
%\frac{q v_d}{\epsilon} \frac{g_0^+ + g_0^-}{2}\frac{\tau_v}{\tau_d}\frac{(-1+\Omega^2\tau_v\tau_d)}{\Omega(1+\Omega^2\tau_v^2)}
%-\frac{q v}{\epsilon}\frac{g_0^+ - g_0^-}{2}\frac{\tau_v}{\tau_s}\frac{1+\Omega^2\tau_s\tau_v}{\Omega(1+\Omega^2\tau_v^2)} \nonumber\\
B &=\frac{q v_d}{\epsilon} \frac{\overline{g_0^+} + \overline{g_0^-}}{2}\left(-\frac{2}{\tau_v}+\frac{1}{\tau_r}\right) \tau_a^2-\frac{q v}{\epsilon}\frac{\overline{g_0^+} - \overline{g_0^-}}{2}\frac{\tau_a^2}{\tau_r}\nonumber\\
%= \frac{q v_d}{\epsilon} \frac{g_0^+ + g_0^-}{2}\frac{\tau_v}{\tau_d}\frac{(\tau_v+\tau_d)}{(1+\Omega^2\tau_v^2)} + 
%+\frac{q v}{\epsilon}\frac{g_0^+ - g_0^-}{2}\frac{\tau_v}{\tau_s} \frac{(\tau_v-\tau_s)}{(1+\Omega^2\tau_v^2)}\nonumber\\
C &=\frac{q v_d}{\epsilon} \frac{\overline{g_0^+} + \overline{g_0^-}}{2} \left(\frac{2}{\tau_v}-\frac{1}{\tau_r}\right) \tau_a^2\nonumber\\
D &= \frac{q v}{\epsilon}\frac{\overline{g_0^+} - \overline{g_0^-}}{2}\frac{\tau_a^2}{\tau_r}.
\end{align}
Since we are considering a delta-function pulse at $t=0$, the influence of the pulse in $V_{\rm int}(t)$ should vanish when $t<0$ and $t\to \infty$.
This imposes
\begin{equation}
V_{\rm int}(0^-) =V_{\rm int}(\infty) =0.
\end{equation}
We notice also that $V_{\rm int}$ is continuous at $t=0$ ($B+C+D=0$), as expected by integrating the left and right sides of Eq.~(\ref{eq:dVdt0}) across $t=0$.
The last two lines in Eq.~(\ref{eq:dVdt0}), which appear to be higher order in $\tau_a$, are in fact important in order to obtain a solution that satisfies $V_{\rm int}(0^+)= V_{\rm int}(\infty)=0$.
We also notice that $A$ is parametrically larger than $B$, $C$ and $D$ in the $\tau_a\ll(\tau_r,\tau_v)$ regime. 
Further assuming that $\tau_r$ is not long compared to $\tau_v$, we can approximate
%the transient part of the voltage at $t>0$ in the short $\tau_a$ approximation reads
\begin{equation}
\label{eq:Vsimp}
V_{\rm int}(t) \simeq \frac{q}{\epsilon} \left(\frac{\overline{g_0^{+}}+\overline{g_0^{-}}}{2} v_d -\frac{\overline{g_0^{+}}-\overline{g_0^{-}}}{2} v\right) \tau_a e^{-t/\tau_v} \sin(t/\tau_a).
\end{equation}
%where we have neglected $B/A$ for short $\tau_a$.
%We have verified that the units of Eq.~(\ref{eq:Vsimp}) are correct. 
%{\bf It would be nice to have an intuitive picture that explains this result.}

Equation~(\ref{eq:Vsimp})  was derived assuming a very short pulse in time. In practice, this would require pulses faster than $\tau_a$, which may imply subpicosecond times for typical magnetic fields required to attain the quantum limit. For such short pulses, one might be concerned that carriers of additional nonchiral Landau levels not included in our theory could be excited significantly.
% (recall that 1 picosecond is roughly associate to excitation energies of a couple of meV).
Thus, if we are interested in restraining the carrier dynamics to only $n=0$ and $n=1$, it is necessary to consider light pulses that are slower than $\tau_a$. 
Hence, for completeness, we will consider a Gaussian light pulse with a nonzero width $t$, i.e. 
\begin{equation}
\overline{g^\chi}(t)= \overline{g_0^\chi}\frac{1}{\sqrt{2\pi} dt} e^{-t^2/(2 dt^2)}
\end{equation}
for a constant $g_0^\chi$, which implies
\begin{equation}
\overline{\tilde{g}^\chi}(\omega) = g_0^\chi e^{-\omega^2 dt^2/2}.
\end{equation}
In the limit $dt\to 0$, we recover the results from the previous paragraphs. Let us now see how those results change when $dt\neq 0$, and possibly $dt>\tau_a$.

It turns out that the solutions for $\overline{\Sigma_1}$ and $\overline{\Delta_0}$ can still be obtained using the residue theorem to a good approximation, provided that we consider $t \gg dt$.
In comparison with the solution for the delta-function pulse, the solutions for the Gaussian pulse result in the following substitutions:
\begin{align}
&e^{-t/\tau_r} \to e^{-t/\tau_r} e^{dt^2/(2 \tau_r^2)}\nonumber\\
&e^{-t/\tau_v} \to e^{-t/\tau_v} e^{dt^2 (-1/(2\tau_a^2) +1/\tau_v^2)}\nonumber\\
&\cos(\Omega t) \to \cos\left(\Omega(t - dt^2/\tau_v)\right)\nonumber\\
&\sin(\Omega t) \to \sin\left(\Omega(t - dt^2/\tau_v)\right).
\end{align}
Consequently, we find that the counterpart of Eq.~(\ref{eq:Vsimp}) becomes
\begin{equation}
V_{\rm int}(t) \simeq \frac{q}{\epsilon} \left(\frac{\overline{g_0^{+}}+\overline{g_0^{-}}}{2} v_d -\frac{\overline{g_0^{+}}-\overline{g_0^{-}}}{2} v\right) \tau_a e^{-t/\tau_v} e^{-dt^2/(2 \tau_a^2)}\sin\left[\frac{t-dt^2/\tau_v}{\tau_a}\right],
\end{equation}
which matches with Eq.~(18) of the main text for $t > dt$ (in the main text, we wrote $t-dt^2/\tau_v\simeq t$ for $t>dt$, as our regime of interest is $dt\ll \tau_v$).

\section{Weak magnetic field regime}
\label{sec:weak_field}

In this section, we adapt our theory to the case of weak magnetic fields, where Landau quantization can be ignored. 
The energy spectrum is then made of two linearly dispersing Weyl cones with a constant group velocity $v$. The two nodes are separated from one another in momentum space.
Like in the main text, we assume that the two Weyl cones are related to one another by an improper symmetry.

We denote the bands as $(n,\chi)$, where $n=c,v$ indicates the conduction or valence band and $\chi$ labels the chirality.
We assume that the Fermi level intersects the valence bands deep enough so that the thermal population of electrons in the conduction band is negligible.
Moreover, we adopt the standard semiconductor convention of describing the carriers as electrons in the conduction band and holes in the valence bands (this differs from the main text, where we used the electron picture to describe the carrier dynamics in both $n=0$ and $n=1$ Landau levels).
Thus, hereafter $\rho_c^\chi$ ($\rho_v^\chi$) denotes the electron (hole) concentration in the conduction (valence) band of chirality $\chi$. 
%Moreover, unlike in the main text, we will represent 
%Let us see what happen in the case of a weak magnetic field applied on a Weyl semimetal. By weak, we mean that the Landau quantisation is completely negligible so now we'll just talk about conduction and valence bands. Consequently, instead of using the Landau level index in the notation ($n, \chi$), we'll use the band type. For example, the conduction band of the Weyl node $"+"$ is denoted by ($c, +$) and the valence band in the Weyl node "$-$" by ($v, -$).

%Moreover, the electron formalism does not stand anymore. We'll use electrons to describe the charge carrier density in conduction bands but we'll use holes to characterise it in valence bands. In fact, considering our linear model for Weyl nodes, without the chiral level ($n=0$) as in the strong field case, the equilibrium electrons density would be infinite in valence bands. The notation will be similar as before, the electron density in the conduction band ($c, \chi$) is denoted $\rho_c^{\chi}$ and the holes density in the valence band ($v, \chi$) is denoted $\rho_v^{\chi}$.
%\\

%In this study, all assumptions made on the light pulse and on the dimensionality of the equations remain valid. So we can still use the one-dimensional van Roosbroeck's equations and, applying the drift-diffusion approach, 
The electric current carried by electrons in the conduction band of chirality $\chi$ can be written as 
\begin{equation}
\label{eq:jc}
    j_c^{\chi} = q\mu_c^\chi \rho_c^{\chi}E + q D_c^\chi\partial_z \rho_c^{\chi},
\end{equation}
where $E$ is the $z-$component of the electric field, $\mu_c^\chi>0$ is the electron mobility and $D_c^\chi>0$ is the diffusion coefficient.
Similarly, the electric current carried by holes in the valence band of chirality $\chi$ can be written as
\begin{equation}
    \label{eq:jv}
    j_v^{\chi} = q\mu_v^\chi \rho_v^{\chi}E - qD_v^\chi \partial_z\rho_v^{\chi},
\end{equation}
where $\mu_v^\chi >0$ is the hole mobility and $D_v^\chi>0$ is the diffusion coefficient.
% ($m^*_h$ is the effective mass of holes in the valence band). In contrast to the ($n=0$)

The use Eqs.~(\ref{eq:jc}) and (\ref{eq:jv}) for Weyl fermions requires some comments. In a simple parabolic electronic band $n$, the drift current can be written as $q\mu_n\rho_n E$, where the mobility $\mu_n$ is approximately independent from the carrier concentration $\rho_n$.
The situation changes for a linearly dispersing Weyl band. 
In this case, while we may insist to write the drift current as $q\mu_n^\chi\rho_n^\chi E$, the mobility can no longer be considered to be approximately independent from the carrier concentration.
For example, in the absence of a magnetic field and at low temperature, the equilibrium hole concentration in the valence band can be written as
\begin{equation}
\rho_v^\chi =\frac{1}{6\pi^2\hbar^3} \frac{\epsilon_F^3}{v^3},
\end{equation}
where $\epsilon_F>0$ is the Fermi energy measured from the Weyl node.
Under the same conditions, the mobility reads 
\begin{equation}
\mu_v^\chi = q v^2 \tau/\epsilon_F,
\end{equation}
 where $\tau$ is the electronic lifetime.
Thus, if $\rho_v$ varies out of equilibrium due to (say) a change in $\epsilon_F$, then so does $\mu_v$.
Concerning the diffusion coefficients $D_n^\chi$, they are related to the conductivities $\sigma_n^\chi=q \mu_n^\chi\rho_n^\chi$ via the Einstein relation. 
At zero magnetic field and low temperature, we obtain $D_n^\chi\simeq v^2 \tau/3$, which can approximated as independent from the carrier concentration.
It is with these qualifications that we write  Eqs.~(\ref{eq:jc}) and (\ref{eq:jv}).

Charge continuity equations now read
%will be slightly different due to the change of sign between electrons and holes, they now read 
\begin{align}
\label{eq:contweak}
    \frac{\partial j_{c}^{\chi}}{\partial z} - q \frac{\partial \rho_{c}^{\chi}}{\partial t} &= q R_c^\chi + q G_c^\chi\nonumber\\
     \frac{\partial j_{v}^{\chi}}{\partial z} + q \frac{\partial \rho_{v}^{\chi}}{\partial t} &= -q R_v^\chi - q G_v^\chi - \chi \frac{q^3}{4\pi^2\hbar^2}E B_0,
\end{align}
where in the low temperature limit that we consider the chiral anomaly affects only the band that crosses the Fermi level (the valence band in our case), and
\begin{equation}
G_c^\chi= G_v^\chi= -\frac{g^{\chi}}{2}.
\end{equation}
Note that the light pulse increases the number of electrons in the conduction band and the number of holes in the valence band.
%In fact, in this case, saying that there is a gain of one electron in the conduction band is exactly saying that there is a gain of one hole in the valence band. 
Concerning the relaxation rate for the excess in the charge, we have
%So for the conduction band of chirality $\chi$, this relaxation rate is  
\begin{align}
\label{eq:relweak}
 &  R_c^\chi= \frac{\rho_c^{\chi} - \rho_{c, \rm eq}^{\chi}}{\tau_r} + \chi \frac{\rho_c^+ - \rho_c^-}{\tau_v}\nonumber\\
&  R_v^\chi= \frac{\rho_c^{\chi} - \rho_{c, \rm eq}^{\chi}}{\tau_r} + \chi \frac{\rho_v^+ - \rho_v^-}{\tau_v}, 
%+ \chi \frac{q^2}{4\pi^2\hbar^2}E B_0,
\end{align}
where $\rho_{n, \rm eq}^{\chi}$ are the equilibrium densities.
%For the valence band, the relaxation rate is
%\begin{equation}
%    \left( \frac{\partial \rho_{v}^{\chi}}{\partial t}\right)_{\text{rel}} = \frac{\rho_c^{\chi} - \rho_{c, eq}^{\chi}}{\tau_r} + \chi \frac{\rho_v^+ - \rho_v^-}{\tau_v} + \chi \frac{q^2}{4\pi^2\hbar^2}EB_0
%\end{equation}
Like in the main text, the total charge is conserved. 

Concerning Poisson's equation, it reads
%In the van Roosbroeck's system, there is the Poisson equation to characterise the electric field, considering the same assumptions as before, it now reads
\begin{equation}
\label{eq:poweak}
\frac{\partial E}{\partial z} = -\frac{q}{\epsilon} \sum_{\chi=\pm1} \left[ (\rho_c^{\chi} - \rho_{c, \rm eq}^{\chi}) - (\rho_v^{\chi} - \rho_{v, \rm eq}^{\chi}) \right],
\end{equation}
where we assume that, in equilibrium, the doping concentration is uniform. 
%where $\rho_{v, eq}^{\chi})$ is the finite hole density in the valence band of chirality $\chi$ at equilibrium. 

\begin{comment}

To summarize, the van Roosbroeck's system of equations becomes
\begin{subequations}
    \begin{align}
        &j_c^{\chi} = q\mu_c \rho_c^{\chi}E + qD_c\partial_z \rho_c^{\chi} \\
        &j_v^{\chi} = q\mu_v \rho_v^{\chi}E - qD_v \partial_z\rho_v^{\chi} \\
        &\frac{\partial j_{c}^{\chi}}{\partial z} - q \frac{\partial
        \rho_{c}^{\chi}}{\partial t} = -\frac{g^{\chi}}{2} + \frac{\rho_c^{\chi} - \rho_{c, eq}^{\chi}}{\tau_r} + \chi \frac{\rho_c^+ - \rho_c^-}{\tau_v} \\
        &\frac{\partial j_v^{\chi}}{\partial z} + q \frac{\partial \rho_{v}^{\chi}}{\partial t} = \frac{g^{\chi}}{2} - \frac{\rho_c^{\chi} - \rho_{c, eq}^{\chi}}{\tau_r} - \chi \frac{\rho_v^+ - \rho_v^-}{\tau_v} - \chi \frac{q^2}{4\pi^2\hbar^2}EB_0\\
        &\frac{\partial E}{\partial z} = -\frac{q}{\epsilon} \sum_{\chi=\pm1} \left[ (\rho_c^{\chi} - \rho_{c, \rm eq}^{\chi}) - (\rho_v^{\chi} - \rho_{v, \rm eq}^{\chi}) \right].
    \end{align}
\end{subequations}

\end{comment}

Eqs.~(\ref{eq:jc}), (\ref{eq:jv}), (\ref{eq:contweak}), (\ref{eq:relweak}) and (\ref{eq:poweak}) form the van Roosbroeck system of equations at weak magnetic field. 
These equations can be linearized in the same way as in App.~\ref{sec:linearized}, except for two changes.
The first change comes from the fact that we must take into consideration the dependence of the mobility on the carrier concentration.
For example, we write
\begin{equation}
\label{eq:parj}
\partial_z(j_n^\chi) = q (\partial_z\mu_n^\chi) \rho_n^\chi E+ q \mu_n^\chi (\partial_z\rho_n^\chi) E+q \mu_n^\chi \rho_n^\chi (\partial_z E)\pm q D_n^\chi \partial^2_z \rho_n^\chi,
\end{equation}
where the $+$ ($-$) sign is for $n=c$ ($n=v$).
In parabolic band systems, the first term in the right hand side of Eq.~(\ref{eq:parj}) would be omitted.
In the present case, we express
\begin{equation}
\frac{\partial\mu_n^\chi}{\partial z} = \frac{\partial \mu_n^\chi}{\partial \rho_n^\chi}\frac{\partial \rho_n^\chi}{\partial z}.
\end{equation}
Thus,
\begin{equation}
\partial_z(j_n^\chi) = q \tilde{\mu}_n (\partial_z\rho_n) E+q \mu_n \rho_n (\partial_z E),
\end{equation}
where
\begin{equation}
\tilde{\mu}_n^\chi = \mu_n^\chi + \rho_n^\chi \frac{\partial \mu_n^\chi}{\partial \rho_n^\chi}
\end{equation}
is a modified mobility.
Upon linearization, we write 
\begin{equation}
q \tilde{\mu}_n^\chi (\partial_z\rho_n) E \simeq q \tilde{\mu}_{n,\rm eq} (\partial_z\rho_n^\chi) E_0,
\end{equation}
where the equilibrium mobility $\tilde{\mu}_{n,\rm eq}$ is the same for both chiralities due to the crystal symmetry relating the two Weyl cones.

The second change comes from the fact that the equilibrium hole concentration in the valence band ($\rho_{v,\rm eq}^\chi$) is not small in our theory, and hence 
\begin{equation}
q\mu_v \rho_{v,\rm eq}^\chi \partial_z E\simeq q\mu_{v,\rm eq} \rho_{v,\rm eq}^\chi \partial_z E_{\rm int}
\end{equation}
will not be neglected.
%we do not neglect the term $q\mu_v \rho_{v,\rm eq}^\chi \partial_z E_{\rm int}$ in $\partial_z j^\chi_c$.

Therefore,  the linearization of van Roosbroeck's equations yields
%About the linearisation of these equations, the approach is exactly the same is in App. B using also the same assumptions. So after this linearisation, the new linearised van Roosbroeck's system is }
    \begin{align}
    \label{eq:vrweak}
    %\label{eq:poissonweak}
        &\partial_z E_{\rm int} = -\frac{q}{\epsilon}\left(\Sigma_c - \Sigma_v  \right) \nonumber\\
        &\left(\tilde{\mu}_{c,\rm eq}E_0\partial_z + D_c \partial_z^2 - \partial_t - \frac{1}{\tau_r} \right)\Sigma_c = -g_s \nonumber\\
        &\left(\tilde{\mu}_{c,\rm eq}E_0\partial_z + D_c \partial_z^2 - \partial_t - \frac{1}{\tau_r} - \frac{2}{\tau_v} \right)\Delta_c = -g_d \nonumber\\
        &\left( -\tilde{\mu}_{v,\rm eq} E_0\partial_z + D_v \partial_z^2 - \partial_t   \right)\Sigma_v = -g_s + \frac{1}{\tau_r}\Sigma_c +\mu_{v,\rm eq} \left(\rho_{v,\rm eq}^+ + \rho_{v,\rm eq}^-\right)\partial_z E_{\rm int}\nonumber\\
        & \left(  -\tilde{\mu}_{v,\rm eq} E_0\partial_z + D_v \partial_z^2 - \partial_t - \frac{2}{\tau_v}\right)\Delta_v = -g_d + \frac{1}{\tau_r}\Delta_c +  \frac{q^2}{2\pi^2\hbar^2}E B_0,
    \end{align}
where $\Sigma_n \equiv \sum_{\chi}(\rho_n^{\chi} - \rho_{n,\rm eq}^{\chi})$ and $\Delta_n \equiv \sum_{\chi}\chi(\rho_n^{\chi} - \rho_{n,\rm eq}^{\chi}) $.
% are the analogues of $\Sigma_n$ and $\Delta_n$ in the strong field case. 

%Now, following the App. C and using these properties about the n-time derivative of the Fourier transform of an integrable function $f(z, t)$ over $\mathbb{R}$ 
%\begin{subequations}
%\begin{align}
%      \partial^n_z \Tilde{f}(k, \omega) &= (-ik)^n \Tilde{f}(k, \omega) \nonumber  \vspace{0.3cm}\\
%   \partial^n_t \Tilde{f}(k, \omega) &= (i\omega)^n \Tilde{f}(k, \omega) \nonumber
%\end{align}
%\end{subequations}
%we can express the van Roosbroeck's equations in the Fourier space which is solvable algebraically :
%\begin{subequations}
%    \begin{align}
%        &ik\tilde{E}_{int} = \frac{q}{\epsilon}\left(\tilde{\Sigma}_c - \tilde{\Sigma}_v  \right) \\
 %       &\left(-\mu_cE_{app}ik  -D_c k^2 - i\omega - \frac{1}{\tau_r} \right)\tilde{\Sigma}_c = -\tilde{g}_s \\
 %       &\left(-\mu_cE_{app}ik - D_c k^2 - i\omega - \frac{1}{\tau_r} - \frac{2}{\tau_v} \right)\tilde{\Delta}_c = -\tilde{g}_d \\
 %       &\left( \mu_vE_{app}ik - D_v k^2 - i\omega   \right)\tilde{\Sigma}_v = -\tilde{g}_s + \frac{1}{\tau_r}\tilde{\Sigma}_c \\
 %       & \left( \mu_vE_{app}ik - D_v k^2 - i\omega - \frac{2}{\tau_v}\right)\tilde{\Delta}_v = -\tilde{g}_d + \frac{1}{\tau_r}\tilde{\Delta}_c + \frac{q^2}{2\pi^2\hbar^2}\tilde{E}B_0
 %   \end{align}
%\end{subequations}

Now, let us study the transient photovoltage using the same approach as in the strong field regime. We reconsider the situation where a Weyl semimetal film of length $L$ along the $z$ direction is placed between two contacts and subjected to a light pulse  centered at time $t=0$. The light pulse acts far enough from the contacts, such that carrier densities at $z=\pm L/2$ remain at their equilibrium values. 
In this condition, Eqs.~(\ref{eq:vrweak}) can be solved in Fourier space, like in Sec. C.
As a result, the following relation is verified for the particle photocurrent:
%So we can still write \eqref{eq:jD} but the particle photocurrent will be slightly different :
\begin{equation}
\label{eq:djweak}
    \delta j = - q v_c\Sigma_c + q v_v\Sigma_v +q \mu_{v,\rm eq} \left(\rho_{v,\rm eq}^+ + \rho_{v,\rm eq}^-\right) E_{\rm int}+ q D_c\partial_z \Sigma_c - qD_v\partial_z \Sigma_v =-\epsilon \partial_t E_{\rm int},
\end{equation}
where $v_c = -\tilde{\mu}_{c,\rm eq} E_0$ is the drift velocity of electrons in the conduction band  and $v_v = \tilde{\mu}_{v,\rm eq} E_0$ is the drift velocity of holes  in the valence band.
Integrating \eqref{eq:djweak} over the sample length and applying the boundary conditions, we obtain
% \eqref{defmoy} and remembering the definition on the electric potential $V = -\overline{E_{int}}$, we obtain
\begin{equation}
\label{eq:potweak}
    \epsilon\partial_t V_{\rm int} = \overline{\delta j} = -qv_c \overline{\Sigma_c} + qv_v \overline{\Sigma_v}-q \mu_{v,\rm eq} \left(\rho_{v,\rm eq}^+ + \rho_{v,\rm eq}^-\right) V_{\rm int},
\end{equation}
where $\overline{\Sigma_c}$ and $\overline{\Sigma_v}$ obey, according to the spatial integration of Eq.~\eqref{eq:vrweak}, 
\begin{subequations}
\label{eq:barvr}
    \begin{align}
        &\left(\partial_t +\frac{1}{\tau_r} \right)\overline{\Sigma_c} = \overline{g_s} \\
        &\partial_t\overline{\Sigma_v} = \overline{g_s} -\frac{1}{\tau_r}\overline{\Sigma_c}.
    \end{align}
\end{subequations}
Considering a delta function light centered at $t=0$, i.e. $\overline{g_s}=  \overline{g_{0s}}  \delta(t)$,  Eq.~(\ref{eq:barvr}) gives
%
% pulse defined in the same way that before \eqref{deflightpulse}, equations \eqref{eq:barvr} can be integrated in the Fourier space using the residue theorem which is still a good approximation if $t \gg dt$ as explained in App. C. We obtain 
\begin{equation}
  %  \overline{\Sigma_c} = \overline{\Sigma_v} = \overline{g_{0s}} \,e^{dt^2 /2\tau_r^2} e^{-t/\tau_r} \,\Theta(t)
    \overline{\Sigma_c} = \overline{\Sigma_v} = \overline{g_{0s}} e^{-t/\tau_r} \,\Theta(t),
\end{equation}
where $\Theta(t)$ is the step function.
The fact that $  \overline{\Sigma_c} = \overline{\Sigma_v}$ can also be obtained from in integration of Poisson's equation with the boundary condition that $E_{\rm int}$ is negligible at the contacts. 
%This result may appear surprising but, considering the integration over the sample of the Poisson equation \eqref{eq:poissonweak}, it leads to the fact that the average internal electric field is zero. So the result becomes consistent with the physical situation.

Consequently, \eqref{eq:potweak} becomes
\begin{equation}
    \partial_t V_{\rm int} = \frac{q}{\epsilon}  \overline{g_{0s}} (v_v - v_c) e^{-t/\tau_r} \Theta(t) - \frac{\sigma_v}{\epsilon} V_{\rm int},
\end{equation}
where $\sigma_v = q \mu_{v,\rm eq} \left(\rho_{v,\rm eq}^+ + \rho_{v,\rm eq}^-\right) $ is the equilibrium conductivity at low temperature and $\epsilon/\sigma_v\equiv\tau_D$ is the dielectric relaxation time for excess charge in the valence band. 
Integrating over time, using $V_{\rm int} (t<0)=0$ and imposing the continuity of $V_{\rm int}$ at $t=0$, we arrive at
\begin{equation}
\label{eq:V_int_weak}
V_{\rm int}(t) = \frac{q}{\epsilon} \overline{g_{0s}} (v_v - v_c) \frac{\tau_D \tau_r}{\tau_D -\tau_r} \left(e^{-t/\tau_D} - e^{-t/\tau_r}\right).
\end{equation}
%which can be integrated over time and we get
%\begin{equation}
%V(t) =\left\{\begin{array}{cc} c e^{-t/\tau_D} +\frac{q}{\epsilon} \overline{g_{0s}} (v_v - v_c) \frac{\tau_r\tau_D}{\tau_r-\tau_D} e^{-t/\tau_r} &\text{ for } t>0\\
%V(0^-) &\text{ for } t<0,
%\end{array}\right.
%\end{equation}
%where $V(0+) = 0$ and $V(0^-) = 0$ due to the light pulse characteristics when $t$ goes to zero or to infinity. So finally, the transient photovoltage of a Weyl semimetal subjected to a weak magnetic field is 
%\begin{equation}
%    V(t) = \frac{q}{\epsilon} \overline{g_{0s}} (v_c - v_v)e^{dt^2 /2\tau_r^2} \tau_re^{-t/\tau_r}
%\end{equation}
%This difference between drift velocities of the conduction and valence band suggests that we must have an asymmetry between these two bands. In fact, if the conduction and valence band are completely symmetric, then electrons and holes would have the same drift velocity. So in this case, there would be no transient photovoltage in the Weyl semimetal.
The fact that photoexcited electrons and holes drift in opposite directions under the action of $E_0$ (i.e. $v_c v_v<0$) is crucial for the development of the transient photovoltage. 
Some limiting regimes of Eq.~(\ref{eq:V_int_weak})  are
\begin{align}
V_{\rm int}(t) & \simeq \frac{q}{\epsilon} \overline{g_{0s}} (v_v - v_c) \tau_D e^{-t/\tau_r} \text{ , for  } \tau_D\ll \tau_r\nonumber\\
V_{\rm int}(t) & \simeq \frac{q}{\epsilon} \overline{g_{0s}} (v_v - v_c) \tau_r e^{-t/\tau_D} \text{ , for  } \tau_D\gg \tau_r.
\end{align}
In sum, the transient photovoltage at weak magnetic fields decays in a nonoscillatory fashion.
This behavior differs qualitatively from the strong magnetic field regime (see main text), where $V_{\rm int}$ oscillates at the plasma frequency due to the chiral anomaly term in the van Roosbroeck equations.
% and has no direct influence from the chiral anomaly. 
At weak fields, the chiral anomaly term in Eq.~(\ref{eq:relweak}) does not enter in the transient photovoltage.
Nevertheless, the chiral anomaly influences $V_{\rm int}$ indirectly, through its participation in the drift velocities. 
It is well-known that, at weak field, chiral anomaly causes an anisotropy of order $B_0^2$ in the conductivity tensor \cite{Son2012}.
Since the drift velocity scales with the conductivity, it is different (by an amount of order $B_0^2$) when the magnetic field is parallel or perpendicular to the applied electric field.
Thus, the much-studied anisotropic magnetoresistance of Weyl semimetals finds a counterpart in the transient photovoltage under a pulsed light.

\end{widetext}

\bibliography{refs_resub}{}
\bibliographystyle{apsrev4-1}

\end{document}